\documentclass[floatfix,aps,preprint,superscriptaddress,nofootinbib,showpacs,amsmath,amssymb,prd,12pt]{revtex4}
\usepackage{graphicx}
\usepackage{epsfig}

\newcommand{\alphas}{\alpha_s}

\begin{document}
\title{Threshold Resummation Effects in
Neutral Higgs Boson Production by Bottom Quark Fusion at the CERN
Large Hadron Collider}
\author{Hua Xing Zhu}
\email{hxzhu@pku.edu.cn}
\author{Chong Sheng Li}
\email{csli@pku.edu.cn}
\author{Jia Jun Zhang}
\email{jiajunzhang@pku.edu.cn}
\author{Hao Zhang}
\email{haozhang.pku@pku.edu.cn}
\author{Zhao Li}
\email{zhli.phy@pku.edu.cn} \affiliation{Department of Physics and
State Key Laboratory of Nuclear Physics and Technology, Peking
University, Beijing 100871, China}


\begin{abstract}
We investigate the QCD effects in the production of neutral Higgs
bosons via bottom quark fusion in both the standard model and the
minimal supersymmetric standard model at the CERN Large Hadron
Collider. We include the next-to-leading order (NLO) QCD corrections
(including supersymmetric QCD) and the threshold resummation
effects. We use the soft-collinear effective theory to resum the
large logarithms near threshold from soft gluon emission. Our
results show that the resummation effects can enhance the total
cross sections by about 5\% compared with the NLO results.
\end{abstract}
\maketitle

\section{INTRODUCTION}
The understanding of Electroweak Symmetry Breaking (EWSB) plays a
key role in current research of high energy physics. In the Standard
Model (SM), a complex scalar doublet is responsible for the
generation of gauge bosons and fermions masses by the Higgs
mechanism.
One neutral Higgs boson ($h$) survives after EWSB, which is the last
elementary particle yet to be found in the SM. Direct searches at
LEP$2$ set a lower bound on the SM Higgs boson mass $m_h>114.4$ GeV
(at 95\%CL)~\cite{Barate:2003sz}, while electroweak precision
measurements prefer a light Higgs boson of $m_h\lesssim 180$
GeV~\cite{Alcaraz:2007ri}.

In the most popular extensions of the SM, e.g., the Minimal
Supersymmetric Standard Model (MSSM), two Higgs doublets are
required in order to preserve supersymmetry (SUSY) and anomaly
cancellation. In the MSSM, the Higgs sector consists of five
physical Higgs bosons: the neutral CP-even ones $h$ and $H$, the
neutral CP-odd one $A$, and the charged ones $H^\pm$. The lightest
one $h$ behaves like the SM one in the decoupling limit
$(M_{A}\gg M_{Z^0})$. Its mass is constrained by a theoretical upper
bound of $M_{h} \lesssim (130-140)$ GeV when taking into account the
radiative corrections~\cite{massh0}. At lowest order, two parameters
are required to describe the MSSM Higgs sector, which are generally
chosen to be $m_A$ and $\tan\beta=v_2/v_1$, the ratio of the two
vacuum expectation values.

For large value of $\tan\beta$, the bottom-Higgs Yukawa coupling can
be considerably enhanced, thus Higgs production associated with
bottom quarks may be quite important in MSSM. There are two
approaches for calculating cross sections involving bottom quarks:
the four flavor number scheme (4FNS) and five flavor number scheme
(5FNS)~\cite{Dawson:2005vi}. In the 4FNS, there are no initial state
bottom quarks. Due to non zero of bottom quark mass, large
logarithms may appear from gluon splitting, hence transverse
momentum ($p_T$) and pseudo rapidity cuts on final state bottom
quarks are needed to eliminate these large logarithms. In the 5FNS,
initial state bottom quarks are treated as massless, and by
introducing a perturbatively defined bottom quark parton
distribution function (PDF), large logarithms are resummed through
the DGLAP evolution equation. Except for $q\bar{q} \rightarrow
b\bar{b}h$, the other relevant production mechanisms depend on the
final state being observed~\cite{Campbell:2004pu}. For inclusive
Higgs production without bottom tagged in final state, the lowest
order process is $gg\rightarrow (b\bar{b})h$ ($b\bar{b}\rightarrow
h$) in 4FNS (5FNS)~\cite{Dicus:1998hs}. However, if at least one
high-$p_T$ $b$ quark is required to be observed, the leading
partonic process is $gg \rightarrow b(\bar{b})h$ ($gb\rightarrow
bh$) in 4FNS (5FNS)~\cite{gbbh}, and if two high-$p_T$ $b$ quarks
are required, the leading subprocess is $gg\rightarrow
b\bar{b}h$~\cite{ggbbh} and can only be calculated in 4FNS. There
are extensive comparison between 4FNS and 5FNS and good agreement
has been found between the two schemes within theoretical
uncertainties~\cite{Campbell:2004pu,comparison}.

In recent years much effort has been made to the precise prediction
of the inclusive cross section for Higgs boson production via
bottom quark fusion, with neither bottom quark detected
$b\bar{b}\rightarrow h$. The next-to-leading order (NLO) QCD
corrections~\cite{Dicus:1998hs,Balazs:1998sb,Maltoni:2003pn} and the
next-to-next-to-leading order (NNLO) QCD
corrections~\cite{Harlander:2003ai} to this process have been
calculated. Also  the SUSY QCD and SUSY electroweak corrections to
this process have been studied~\cite{Dittmaier:2006cz}.

When the hard scattering process involves two very different scale,
the fixed order perturbation expansion contains large logarithms of
scale ratio. These terms might spoil the reliability of the
perturbation expansion and need to be resummed to all
orders~\cite{Sterman:1986aj,Catani:1989ne}. In general, the large
logarithms mentioned above can appear when the Higgs boson in the
final state has small transverse momentum or is produced near
threshold. The transverse momentum resummation were calculated in
Refs.~\cite{Field:2004nc,Belyaev:2005bs}, and the threshold
resummation effects are
considered~\cite{Ravindran:2006cg,Kidonakis:2007ww} with the
conventional method~\cite{Sterman:1986aj,Catani:1989ne}, where
partial next-next-next-to-leading-order (NNNLO) results are obtained
by expanding the resummed cross sections. In this paper, we will
further study the complete next-to-leading-logarthmic (NLL)
threshold resummation effects on the production cross sections using
soft-collinear effective theory (SCET)~\cite{scet} in both the SM
and MSSM at the CERN Large Hadron Collider (LHC).

The paper is organized as follows: In Sec.~II and III we present the
analytic results at fixed order. In Sec.~IV we use SCET to derive
the resummed formula for the cross sections. In Sec.~V the
numerical results are presented and discussed. Sec.~VI contains a
brief summary and conclusions.

\section{THE LEADING ORDER RESULTS}
We consider the inclusive process $A(p_a)+B(p_b)\rightarrow H_i(q)+X$, where A
and B are the incoming hadrons with momenta $p_a$ and $p_b$,
$H_i=\{h,H,A\}$ are neutral CP-even or CP-odd Higgs bosons, with
momentum $q$, and X is arbitrary hadronic state.

At hadron colliders the total cross sections can be factorized into
the convolution of the partonic cross sections with appropriate
PDFs:
\begin{equation}
\sigma(M^2_i,s)=\sum_{a,b}\int^1_\tau~dx_a\int^1_{\tau/x_a}~dx_b
~f_{a/A}(x_a,\mu_f)f_{b/B}(x_b,\mu_f)
\hat\sigma_{ab}(z,M^2_i;\alphas(\mu^2_r),\mu^2_r,\mu^2_f),
\end{equation}
where $\hat\sigma_{ab}$ is the cross section for the partonic
subprocess $a(\hat{p}_a)+b(\hat{p}_b)\rightarrow H_i(q)+X$,
$\hat{p}_a$ and $\hat{p}_b$ are the momentum of the incoming partons
$a,b$. The momentum fractions $x_a$ and $x_b$ are defined by
$x_{a(b)}=\hat{p}_{a(b)}/p_{a(b)}$, $M_i$ is the mass of the
final state Higgs boson, $\tau=M^2_i/s$, and the scaling
variable $z=M^2_i/\hat{s}$, where $\hat{s}=(\hat{p}_a+\hat{p}_b)^2$.
 $f_{p/H}(x,\mu_f)$ is the parton
distribution function which describes the probability of finding a
parton $p$ with momentum fraction $x$ inside the hadron $H$ at
factorization scale $\mu_f$. The sum is over all possible initial
partons.

\begin{figure}[htp]
 \centering
 \includegraphics[width=0.25\textwidth]{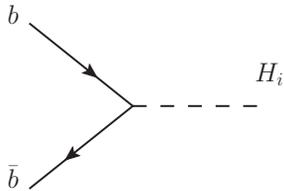}
 \caption{Leading-order Feynman diagrams for $b\bar{b}\rightarrow H_{i}$}
 \label{fig:LO}
\end{figure}

The leading-order (LO) Feynman diagrams is shown in
Fig.~\ref{fig:LO}, and its LO amplitude in $n=4-2\epsilon$ dimension
is
\begin{equation}
i \mathcal{M}_{B}=\mu^{\epsilon}_r Y_b \kappa_i \bar{v}(p_b)
\Gamma^{(i)}u(p_a),
\end{equation}
where $Y_b\equiv -i g m_b/(2 m_W)$ is the SM coupling of Higgs boson
to bottom quark, and $\mu_r$ is a mass parameter introduced to keep
the coupling constant $g$ dimensionless. Furthermore, we have
$\Gamma^{(h,H)}=1$ for scalar ($h,H$) production and
$\Gamma^{(A)}=\gamma^5$ for pseudoscalar ($A$) production. The
explicit expressions for $\kappa_i$ are:
\begin{eqnarray*}
\kappa_{h}=-\frac{\sin\alpha}{\cos\beta}, \ \ \ \ \
\kappa_{H}=\frac{\cos\alpha}{\cos\beta}, \ \ \ \ \
\kappa_{A}=-i\tan\beta.
\end{eqnarray*}
Here $\alpha$ is the mixing angle between the weak and the mass
eigenstates of the neutral CP-even Higgs boson sector.

The LO partonic cross sections are given by
\begin{equation}
\hat{\sigma}_B = \frac{1}{2\hat{s}}\int~\overline{\sum}
|\mathcal{M}_{B}|^2 dPS^{(1)} =\hat{\sigma}_0 \delta (1-z),
\end{equation}
where $\overline{\sum}$ indicates the summation over final states
and the average over initial states, $\int dPS^{(1)}$ represents the
phase space integration and $\hat{\sigma}_0=\mu^{2\epsilon}_r \pi|Y_b
\kappa_i|^2 /(6\hat{s})$.

\section{NEXT-TO-LEADING ORDER CALCULATIONS}

The NLO QCD and SUSY QCD correctons to this process have been
studied
in~\cite{Dabelstein:1995js,Dicus:1998hs,Maltoni:2003pn,Harlander:2003ai,Dittmaier:2006cz},
but we recalculate it here to check the relevant results in the
previous literatures and make our paper self-contained. At NLO, the
QCD and SUSY QCD corrections consist of the following contributions:
the exchange of virtual gluon or gluino and the corresponding
renormalization counterterms, the real gluon emission subprocesses,
the gluon initiated subprocesses, and the contributions of
Altarelli-Parisi(A-P) splitting functions. In the following, we will
calculate these contributions separately. We use dimensional
regularization (DREG) in $n=4-2 \epsilon$ dimensions to regulate all
divergences, and adopt $\overline{\rm MS}$ renormalization and
factorization scheme to remove the ultraviolet (UV) and infrared
(IR) (including soft and collinear) divergences of QCD corrections,
while SUSY QCD corrections are renormalized in on-shell
scheme~\cite{onmass}.

\subsection{Virtual corrections}

The amplitude corresponding to virtual gluons exchange is given by
\begin{equation}
 \mathcal{M}^{\rm QCD}_{V}=\mathcal{M}_{B}\frac{\alpha_s}{4\pi}
C_F\left( \frac{4\pi\mu^2_r}{M^2_i}\right)^\epsilon
\frac{\Gamma(1-\epsilon)}{\Gamma(1-2\epsilon)} \left(
-\frac{2}{\epsilon^2_{\rm IR}}+\frac{3}{\epsilon_{\rm
UV}}-\frac{3}{\epsilon_{\rm IR}}+\frac{2}{3}\pi^2-2\right),
\end{equation}
where $C_F=4/3$ is the $SU(3)$ color factor. The above amplitude
contains both UV and IR divergence. The renormalized QCD amplitude
can be written as
\begin{equation}
\widetilde{\mathcal{M}}^{\rm QCD}_{V} =
  \mathcal{M}^{\rm QCD}_{V}+ \frac{\delta m^{\rm QCD}_b}{m_b}
  \mathcal{M}_{B},
\end{equation}
with~\cite{Braaten:1980yq}
\begin{equation}
\frac{\delta m^{\rm QCD}_b}{m_b}=-C_F\frac{\alpha_s}
{4\pi}\frac{(4\pi)^\epsilon\Gamma(1-\epsilon)}
{\Gamma(1-2\epsilon)}\frac{3}{\epsilon_{\rm UV}}.
\end{equation}

The amplitude corresponding to virtual gluino and squark exchange is
given by
\begin{equation}
 \mathcal{M}^{\rm SUSY}_{V}=\mathcal{M}_B F^{(i)}.
\end{equation}
Here $F^{(i)}$ $(i=h,H,A)$ are the SUSY form factors:
\begin{eqnarray}
F^{(h,H)}&=& \frac{i}{Y_b \kappa_{h,H}}
\frac{\alpha_s}{4\pi}C_F m_{\tilde{g}}\left[\sin 2\theta_{\tilde{b}}
\left(G^{(h,H)}_{11}C_0(1,1)-G^{(h,H)}_{22}C_0(2,2)\right)
\right.
\nonumber
\\
&&\left. +\cos 2\theta_{\tilde{b}} \left(
G^{(h,H)}_{12}C_0(1,2)+G^{(h,H)}_{21}C_0(2,1)\right)\right],
\end{eqnarray}
\begin{equation}
 F^{(A)}= \frac{i}{Y_b \kappa_{A}}
\frac{\alpha_s}{4\pi}C_F m_{\tilde{g}}
\left(G^{(A)}_{12}C_0(1,2)-G^{(A)}_{21}C_0(2,1)\right),
\end{equation}
where $C_0(i,j)$ are the usual Passarino-Veltman three-point
function\cite{denner}
\begin{equation}
 C_0(i,j)\equiv C_0(m^2_b,\hat{s},m^2_b,m^2_{\tilde{g}},
m^2_{\tilde{b}_i},m^2_{\tilde{b}_j}).
\end{equation}
$iG^{(i)}_{lm}$ are the couplings between Higgs boson and sbottom
mass eigenstates, which are given in the appendix.
$m_{\tilde{b}_{1,2}}$ are the sbottom masses, $m_{\tilde{g}}$ is the
gluino mass, and $R^{\tilde b}$ is a $2 \times 2$ matrix defined to
rotate the sbottom current eigenstates into the mass eigenstates:
\begin{equation}
\left(\begin{array}{c} \tilde{b}_1 \\ \tilde{b}_2 \end{array}
\right)= R^{\tilde{b}}\left(\begin{array}{c} \tilde{b}_L \\
\tilde{b}_R \end{array} \right), \ \ \ \ \
R^{\tilde{b}}=\left(\begin{array}{cc} \cos\theta_{\tilde{b}} &
\sin\theta_{\tilde{b}} \\ -\sin\theta_{\tilde{b}} &
\cos\theta_{\tilde{b}}
\end{array} \right),
\end{equation}
with $0 \leq \theta_{\tilde{b}} < \pi$ by convention.
Correspondingly, the mass eigenvalues $m_{\tilde{b}_1}$ and
$m_{\tilde{b}_2}$ (with $m_{\tilde{b}_1}\leq m_{\tilde{b}_2}$) are
given by
\begin{eqnarray}
\left(\begin{array}{cc} m_{\tilde{b}_1}^2 & 0 \\ 0 &
m_{\tilde{b}_2}^2 \end{array} \right)=R^{\tilde{b}}
M_{\tilde{b}}^2 (R^{\tilde{b}})^\dag, \ \ \ \ \
M_{\tilde{b}}^2=\left(\begin{array}{cc} m_{\tilde{b}_L}^2 & X_bm_b
\\ X_bm_b & m_{\tilde{b}_R}^2 \end{array} \right),
\end{eqnarray}
with
\begin{eqnarray}
m^2_{\tilde{b}_L} &=& M^2_{\tilde{Q}} +m_b^2 +m_Z^2\cos2\beta
C_{bL}, \nonumber
\\
m^2_{\tilde{b}_R} &=& M^2_{\tilde{D}} +m_b^2 -m_Z^2\cos2\beta
C_{bR}, \nonumber
\\
X_b &=& A_b -\mu\tan\beta.
\end{eqnarray}
where $C_{bL}=-1/2+\sin^2\theta_W/3$, $C_{bR}=\sin^2\theta_W/3$, and
$M_{\tilde{b}}^2$ is the sbottom mass matrix.
$M_{\tilde{Q},\tilde{D}}$ are soft SUSY-breaking parameters, $A_b$
is the trilinear Higgs-sbottom coupling, and $\mu$ is the Higgsino
mass parameter.

The renormalized SUSY QCD amplitude is
\begin{equation}
\widetilde{\mathcal{M}}^{\rm SUSY}_{V} =
  \mathcal{M}^{\rm SUSY}_{V}+ \left( \frac
{\delta m^{\rm SUSY}_b}{m_b}+\frac{1}{2}(\delta Z_{bL} +\delta
Z_{bR})\right)\mathcal{M}_{B},
\end{equation}
and the renormalization constants in the on-shell scheme are fixed
to be
\begin{eqnarray}
&& \frac{\delta m^{\rm SUSY}_b}{m_b} =  - \frac{\alpha_s}{4\pi} C_F
\left\{\sum_{i=1}^2 \bigg[B_1
-\frac{m_{\tilde{g}}}{m_b} \sin2\theta_{\tilde{b}} (-1)^i
B_0\bigg](m_b^2,m_{\tilde{g}}^2,m_{\tilde{b}_i}^2)\right\},\nonumber
\\
&& \delta Z_{bL}= \frac{\alpha_s}{2\pi}C_F\sum_{i=1}^2(R^{\tilde
b}_{i1})^2 B_1(m^2_b,m_{\tilde{g}}^2,m_{\tilde{b}_i}^2),\nonumber
\\
&& \delta Z_{bR}= \frac{\alpha_s}{2\pi}C_F\sum_{i=1}^2(R^{\tilde
b}_{i2})^2 B_1(m^2_b,m_{\tilde{g}}^2,m_{\tilde{b}_i}^2),\nonumber
\end{eqnarray}
where $B_{0,1}$ are the two-point integrals~\cite{denner}.

After adding the counterterms, the UV divergences in
$\widetilde{\mathcal{M}}^{\rm QCD}_{V}+\widetilde{\mathcal{M}}^{\rm
SUSY}_{V}$ are canceled, but the IR divergent terms still persist.
The partonic subprocess cross section is
\begin{equation}
\label{NLOV}
\begin{split}
 \hat{\sigma}_{V}=
&\hat{\sigma}_{0}\delta(1-z)
\left\{
1 +
\left[
 C_F\frac{\alpha_s}{2\pi}\left(\frac{4\pi\mu^2_r}{M^2_i}\right)^\epsilon
\frac{\Gamma(1-\epsilon)}
{\Gamma(1-2\epsilon)}
\right.
\right.
\\
&\left. \left. \times\left(-\frac{2}{\epsilon^2_{\rm
IR}}+\frac{3}{\epsilon_{\rm UV}}-\frac{3}{\epsilon_{\rm IR}}
+\frac{2}{3}\pi^2-2\right) +2\frac{\delta m^{\rm QCD}_b}{m_b}
\right] +\Delta^{\rm SUSY}\right\},
\end{split}
\end{equation}
where $\Delta^{\rm SUSY}$ is the contribution from SUSY QCD corrections
only, which is free of divergences
\begin{equation}
 \Delta^{\rm SUSY}=2\frac{\delta m^{\rm SUSY}_b}{m_b}
+\delta Z_{bL}+\delta Z_{bR}
+2F^{(i)}.
\end{equation}

\subsection{Real gluon emission and gluon initiated subprocesses}

The partonic cross section of real gluon bremsstrahlung are
\begin{equation}
 \begin{split}
  \hat{\sigma}_R=&\hat{\sigma}_0
C_F\frac{\alpha_s}{2\pi}\left(\frac{4\pi\mu^2_r}{M^2_i}\right)^\epsilon
\frac{\Gamma(1-\epsilon)}
{\Gamma(1-2\epsilon)}
    \left\{ \frac{2}{\epsilon^2_{\rm IR}}\delta (1-z)
    -\frac{2}{\epsilon_{\rm IR}}\frac{1+z^2}{(1-z)_+}
\right.
\\
&\left.
    +4(1+z^2)\left[ \frac{\ln(1-z)}{1-z}\right]_+
    -2\left(\frac{1+z^2}{1-z}\right)\ln z+2(1-z)\right\}
 \end{split}
 \label{sigr}
\end{equation}
The ``plus'' function in Eq.~(\ref{sigr}) is defined as
\begin{equation}
 \int^1_0~dz~f_+(z)g(z)=\int^1_0~dz~f(z)\left( g(z)
-g(1)\right)
\end{equation}
where $g(z)$ is any well-behaved function in the region
$0\leq z\leq 1$.

Combining the contributions of the LO result, the virtual
corrections and the real gluon bremsstrahlung, we obtain
the bare NLO partonic cross section:
\begin{eqnarray}
\label{vpr}
\hat{\sigma}^{\rm bare}_{b\bar{b}}&=&
 \hat{\sigma}_{V}+\hat{\sigma}_{R}
\nonumber
\\
 &=&
\hat{\sigma}_{0}
\left\lbrace  (1+\Delta^{\rm SUSY})\delta(1-z)
+C_F\frac{\alpha_s}{2\pi}
\left(4\pi\right)^\epsilon
\frac{\Gamma(1-\epsilon)}
{\Gamma(1-2\epsilon)}\right.
\nonumber
\\
&&\times\left[ -\frac{2}{\epsilon_{\rm IR}} \left(
\frac{1+z^2}{(1-z)_+}+\frac{3}{2}\delta(1-z)\right)
+\delta(1-z)\left(\frac{2}{3}\pi^2-2\right)+2(1-z) \right.
\nonumber
\\
&&\left.\left.
-2\ln\frac{\mu^2_r}{M^2_i}\frac{1+z^2}{(1-z)_+}
+4(1+z^2)\left(\frac{\ln(1-z)}{1-z}\right)_+
-2\left(\frac{1+z^2}{1-z}\right)\ln z
\right]\right\rbrace
\end{eqnarray}
Now the soft divergences coming from virtual gluons and
bremsstrahlung contributions have canceled exactly according to the
Bloch-Nordsieck theorem~\cite{Bloch:1937pw}.
The remaining divergences are collinear.

In addition to the real gluon bremsstrahlung subprocess, there are
also contributions from the gluon initiated processes, which can be
written as
\begin{equation}
 \begin{split}
\label{init}
 \hat{\sigma}^{\rm bare}_{bg}=
& \hat{\sigma}_0 \frac{\alpha_s}{2\pi} (4\pi)^\epsilon \left\{
\frac{1}{2}\left(z^2+(1-z)^2\right) \left[ -\frac{1}{\epsilon_{\rm
IR}} \frac{\Gamma(1-\epsilon)}{\Gamma(1-2\epsilon)} \right. \right.
\\
&
\left.
\left.
+\ln\left(
\frac{M^2_i}{\mu^2_r}\frac{(1-z)^2}{z}\right)\right]
+\frac{1}{4}(1-z)(7z-3)\right\}
\end{split}
\end{equation}

The bare partonic cross sections in Eqs.~(\ref{vpr}) and
(\ref{init}), which contain the collinear singularities
generated by the radiation of gluons and massless quarks, have a
universal structure, and can be factorized into the following form
to all orders of perturbation theory:
\begin{eqnarray}
\hat \sigma_{ab}^{\rm bare}(z,1/\epsilon_{\rm IR}) =\sum_{c,d}
\Gamma_{ca}(z,\mu_f,1/\epsilon_{\rm IR}) \otimes
\Gamma_{db}(z,\mu_f,1/\epsilon_{\rm IR}) \otimes
\hat{\sigma}_{cd}(z,\mu_f)\,,
\end{eqnarray}
where $\mu_f$ is the factorization scale and $\otimes$ is the
convolution symbol defined as
\begin{eqnarray}
f(z)\otimes g(z)= \int_z^1~\frac{dy}{y} f(y)\,g\left(\frac{z}{y}\right).
\end{eqnarray}
The universal splitting functions
$\Gamma_{cd}(z,\mu_f,1/\epsilon_{\rm IR})$ represent the probability
of finding a parton $c$ with fraction $z$ of the longitudinal
momentum inside the parent parton $d$ at the scale $\mu_f$. They
contain the collinear divergences, and can be absorbed into the
redefinition of the PDF according to mass factorization~\cite{altarelli}.
Adopting the ${\overline {\rm MS}}$
mass factorization scheme, we have to $\mathcal{O}(\alpha_s)$
\begin{eqnarray}
\Gamma_{cd}(z,\mu_f,1/\epsilon_{\rm IR}) &=& \delta_{c d} \delta(1-z)-
\frac{1}{\epsilon_{\rm IR}} \frac{\alpha_s}{2 \pi}
\frac{\Gamma(1-\epsilon)} {\Gamma(1-2\epsilon)}
\left(\frac{4\pi\mu_r^2} {\mu_f^2}\right)^ {\epsilon}
P_{cd}^{(0)}(z),
\end{eqnarray}
where $P_{cd}^{(0)}(z)$ are the leading order Altarelli-Parisi
splitting functions~\cite{Altarelli:1977zs}:
\begin{eqnarray}
P_{qq}^{(0)}(z) &=& \frac{4}{3} \left[\frac{1+z^2}{(1-z)_+}
+\frac{3}{2} \delta(1-z) \right], \nonumber
\\[2ex]
P_{qg}^{(0)}(z) &=&P_{\bar{q}g}^{(0)}(z)= \frac{1}{2}\left[ (1-z)^2 +z^2\right].
\end{eqnarray}

After absorbing the splitting functions
$\Gamma_{cd}(z,\mu_f,1/\epsilon_{\rm IR})$ into the redefinition of
the PDFs through the mass factorization, we derive the
hard scattering cross sections $\hat{\sigma}_{ab}(z,\mu_f)$,
which are free of collinear divergences, and depend on the scale
$\mu_f$:
\begin{eqnarray}
 \hat{\sigma}_{b\bar{b}}(z,\mu_r,\mu_f) &=&
\hat{\sigma}_0
\left\lbrace  (1+
\Delta^{\rm SUSY})\delta(1-z)
+\frac{\alpha_s}{2\pi}
\left[2P^{(0)}_{qq}(z)\ln\frac{M^2_i}{\mu^2_f}
\right. \right.
\nonumber
\\
&&
+C_F\delta(1-z)
\left(3\ln\frac{\mu^2_r}{M^2_i}+\frac{2}{3}\pi^2-2\right)
+4C_F(1+z^2)\left(\frac{\ln(1-z)}{1-z}\right)_+
\nonumber
\\
&&
\left.\left.
-2C_F\frac{1+z^2}{1-z}\ln z+2 C_F(1-z)
\right]\right\rbrace
\\
\hat{\sigma}_{bg}(z,\mu_r,\mu_f)
 &=&
\hat{\sigma}_0 \frac{\alpha_s}{2\pi}
\left[ P^{(0)}_{qg}(z)\ln\left(
\frac{M^2_i}{\mu^2_f}\frac{(1-z)^2}{z}\right)
+\frac{1}{4}(1-z)(7z-3)\right],
\nonumber
\\
\hat{\sigma}_{\bar{b}g}(z,\mu_r,\mu_f)
 &=&\hat{\sigma}_{bg}(z,\mu_r,\mu_f).
\label{nlo}
\end{eqnarray}

Finally, we combine these finite $\hat{\sigma}_{ab}(z,
\mu_r,\mu_f)$ with the appropriate partonic distribution
function to arrive at the NLO cross sections:
\begin{eqnarray}
\sigma^{\rm NLO}&=&\int\,dx_adx_b
 \bigg\{
\bigg[
f_{b/A}(x_a,\mu_f)f_{\bar{b}/B}(x_b,\mu_f)+(x_a\leftrightarrow x_b)
\bigg] \hat{\sigma}_{b\bar{b}}(z,\mu_r,\mu_f)
\nonumber
\\
&&
+\bigg[
f_{b/A}(x_a,\mu_f)f_{g/B}(x_b,\mu_f)
+(x_a\leftrightarrow x_b)
\bigg] \hat{\sigma}_{bg}(z,\mu_r,\mu_f)
\nonumber
\\
&&
+\bigg[
f_{\bar{b}/A}(x_a,\mu_f)f_{g/B}(x_b,\mu_f)
+(x_a\leftrightarrow x_b)
\bigg] \hat{\sigma}_{\bar{b}g}(z,\mu_r,\mu_f)
\bigg\}.
\end{eqnarray}
This result have been obtained before and our result agrees with
those in
Refs.~\cite{Dabelstein:1995js,Dicus:1998hs,Maltoni:2003pn,Harlander:2003ai,Dittmaier:2006cz}.

\section{THRESHOLD RESUMMATION}
The NLO results contain terms like $[\ln(1-z)/(1-z)]_+$ and
$1/(1-z)_+$, which are large near the ``partonic threshold region''
$z\rightarrow 1$. Physically, these singular terms represent a class
of large logarithms of scale ratios, which come from the incomplete
cancellation between real gluon emission and virtual gluon
corrections. These logarithms can be systematically resummed to all
orders by solving the certain evolution equations in Mellin moment
space~\cite{Sterman:1986aj,Catani:1989ne}. One drawback in the
traditional resummation formalism is that the separation of the
contributions from the different scales is not obvious, and some
ingredient in the resummed exponent is not easily identified with a
field-theoretical object. In SCET, the resummation procedure has a
more transparent meaning. Once the factorization properties is
established, a soft scale of interest is separated from the
underlying hard scale in the framework of effective theory. By
evolving from the hard scale to the soft scale through the
renormalization group (RG) equation, the large logarithms can be
resummed to all orders. In this approach, all the ingredients needed
have a clear effective field theory interpretation. In fact,
resummation in SCET have been carried out in deep-inelastic
scattering~\cite{Phys.Rev.D68.114019,DIS}, Drell-Yan
production~\cite{drellyan,Becher:2007ty}, Higgs
production~\cite{Higgs}, thrust rate in $e^+ e^-$
annihilation~\cite{thrust} and heavy colored particle
production~\cite{colorparticle}.

The starting point in effective theory approach to threshold
resummation is the factorization formula for hadronic cross
section~\cite{Collins:1989gx,Bauer:2002nz,Becher:2007ty},
\begin{eqnarray}
\label{factori}
 \sigma(M^2_i,s)&=&\sigma_0\sum_{a,b}\int^1_\tau~\frac{dx_a}{x_a}
\int^1_{\tau/x_a}~\frac{dx_b}{x_b}
~f_{a/A}(x_a,\mu_f)f_{b/B}(x_b,\mu_f)
\nonumber
\\
&&
 \times |C_V(M^2_i,\mu_f)|^2S(\sqrt{\hat{s}}(1-z),M^2_i,\mu_f),
\end{eqnarray}
where $\sigma_0=\pi |Y_b \kappa_i|^2/(6s)$ is the LO total cross section,
$C_V$ is the Wilson coefficient of operator in SCET and $S$ is
soft function. The convolution formula in Eq.~(\ref{factori}) can be
further transformed into product formalism with Mellin
transformation
\begin{eqnarray}
\label{mellinfac}
 \sigma_N(M^2_i,s)&=&\int^1_0\, d\tau\, \tau^{N-1} \sigma(M^2_i,s)
\nonumber
\\
&=& \sigma_0\sum_{a,b}f^N_{a/A}(x_a,\mu_f)
f^N_{b/B}(x_b,\mu_f)|C_V(M^2_i,\mu_f)|^2 S_N(M^2_i,\mu_f),
\end{eqnarray}

As mentioned above, Eq.~(\ref{factori}) and (\ref{mellinfac})
contain large logarithms near partonic threshold, which need to be
resummed to all orders. In the following, we will derive the
evolution equations for the hard matching coefficient $C_V$ and soft
function $S$, respectively, in order to resum these large
logarithms. Before proceeding, it should be pointed out that in
principle, the threshold resummation for Higgs production through
$b\bar{b}$ fusion can be obtained from similar results for Drell-Yan
production~\cite{drellyan,Becher:2007ty}, by replacing the hard
matching coefficient $C_V$ with Eq.~(\ref{cv}). This is due to the
fact that the IR divergences in our case do not depend on the
explicit structure of the vertex, and SUSY QCD corrections do not
give rise to new IR divergences. Nevertheless we present our full
results below.

In the full theory, the neutral Higgs boson $H_i$ production  via
bottom quark fusion is described by the Yukawa coupling
\begin{equation}
 J(x)=Y_b\kappa_i\bar{\psi}(x)\Gamma^{(i)}\psi(x),
\end{equation}
where $\psi$ denotes the quark field coupled with the Higgs boson.
In SCET this coupling can be written as an effective operator
\begin{equation}
 \mathcal{J} = C_{V}(M^2,\mu)Y_b\kappa_i\bar{\xi}_{\bar{n}}
W_{\bar{n}}Y_{\bar{n}}\Gamma^{(i)}Y^\dagger_nW^\dagger_n\xi_n,
\end{equation}
where $\xi_{n(\bar{n})}$ is the hard-collinear (anti-hard-collinear)
bottom quark field and $W_{n(\bar{n})}$ ($Y_{n(\bar{n})}$) denotes
the usual collinear (soft) Wilson lines which are required to
ensure collinear (soft) gauge invariance. $n$
and $\bar{n}$ are two light-cone vectors satisfying
$n^2=\bar{n}^2=0$ and $n\cdot \bar{n}=2$. $C_V(M^2_i,\mu)$
 is the hard matching coefficient which comes from integrating out
hard modes in matching from QCD to SCET. At the tree level we have
$C_V=1$. We can determine the $\mathcal{O}(\alpha_s)$ matching
coefficient by evaluating the difference of on-shell matrix elements
of operators in full theory and SCET. In the dimension
regularization, the facts that IR structure of the full theory and
SCET is identical and the on-shell integrals are scaleless and
vanish in SCET imply that the UV divergences of SCET is just the
negative of the IR divergences of the full theory. Furthermore, the
hard matching coefficient is simply the finite part of the full
theory virtual amplitudes. From the $\mathcal{O}(\alpha_s)$ virtual
corrections Eq.~(\ref{NLOV}) we can obtain the NLO matching
coefficient
\begin{equation}
 C_V(M^2_i,\mu)=1+\frac{\alpha_s}{4\pi}
C_F\left[-\ln^2\left(\frac{\mu^2}{M^2_i}\right) +\frac{7}{6}\pi^2-2
\right]+\frac{1}{2}\Delta^{\rm SUSY}, \label{cv}
\end{equation}
where, for simplicity, we redefined the 't Hooft mass as
$\mu^2\rightarrow \mu^2 e^{\gamma_E}/(4\pi)$ and $\gamma_E$ is the
Euler constant. The RG equation of the above matching coefficient
can be found from the UV divergences in the effective theory,

\begin{equation}
\label{good}
 \frac{d}{d\ln\mu}C_V(M^2_i,\mu)=
\gamma_1 (\mu) C_V(M^2_i,\mu),
\end{equation}
with
\begin{equation}
 \gamma_1 (\mu)=\sum^{\infty}_{n=1}\left[\left(\frac{\alpha_s}{\pi}\right)^n
\left(A^{(n)}_1\ln\frac{M^2_i}{\mu^2}+A^{(n)}_0\right)\right],
\end{equation}
where $\gamma_1(\mu)$ is the anomalous dimension of $C_V$, and
$A^{(n)}_1$ is the well known cusp anomalous dimension
\cite{Korchemsky:1987wg,Korchemskaya:1992je}, originates
from the $1/\epsilon^2$ poles in the UV divergences of the
effective theory. Note that
the anomalous dimension $\gamma_1(\mu)$ itself contains a
$\ln(M^2/\mu^2)$ term, which leads to Sudakov double logarithms
evolution, while the $A^{(n)}_0$ term leads to single
logarithms evolution. We can extract $A_1$ and $A_0$ at the LO from
$\mathcal{O}(\alpha_s)$ virtual corrections
\begin{equation}
 A^{(1)}_1=C_F,\qquad A^{(1)}_0=-\frac{3}{2}C_F.
\end{equation}
In order to reach the NLL accuracy, we need the two loop expression
of $A_1$~\cite{Korchemskaya:1992je}
\begin{equation}
 A_1^{(2)} = \frac{1}{2}C_F \left[ C_A \left(
      \frac{67}{18} - \frac{\pi^2}{6} \right) - \frac{5}{9}n_f \right],
\end{equation}
where $C_A=3$ and $n_f=5$.

The soft function $S$, defined as the closed Wilson loop formed
from the product of the soft Wilson lines in the two
currents~\cite{Becher:2007ty}, describes the real gluon emission
and virtual gluon exchange in the soft limit. At the NLO
it is given by~\cite{Korchemsky:1993uz,Belitsky:1998tc}:
\begin{eqnarray}
 S(\sqrt{\hat{s}}(1-z),M^2_i,\mu_s)
&=&
\delta(1-z)+\frac{\alpha_s(\mu_s)}{2\pi}C_F \left[
\delta(1-z)
\left(\ln^2\frac{\mu^2_s}{M^2_i}-\frac{\pi^2}{2}\right)
\right.
\nonumber
\\
&&
\left.
+8\left(\frac{\ln (1-z)}{1-z}\right)_+
-\frac{4}{(1-z)_+}\ln\frac{\mu^2_s}{M^2_i}
\right].
\end{eqnarray}
In moment space the soft function can be written as
\begin{equation}
  S_N(M^2_i,\mu_s) =
1+\frac{\alpha_s(\mu_s)}{2\pi}C_F
\left(4\ln^2\frac{\bar{N}\mu_s}{M_i}+\frac{1}{6}\pi^2
\right),
\label{softf}
\end{equation}
where $\bar{N}=Ne^{\gamma_E}$. It is manifest in Eq.~(\ref{softf})
that the contribution from the logarithms $\ln \bar{N}$ can be
eliminated by choosing the scale $\mu_s=\mu_I\sim M_i/\bar{N}$. The
same scale choice is also adopted implicitly in the traditional
approach. However, such scale choice should be taken with caution,
it will leads to a Landau pole at $\bar{N}\sim M_i/\Lambda_{\rm
QCD}$ when $\bar{N}$ becomes large. Such spurious Landau pole
singularities is avoided in the SCET approach by assuming
$M_i/\bar{N} \gg \Lambda_{\rm QCD}$~\cite{Phys.Rev.D68.114019}. In
other words, the soft function in Eq.~(\ref{softf}) is only
applicable at a perturbative calculable scale. In principle,
nonperturbative effects might be important and a modeling of the
soft function at nonperturbative scale is then
needed~\cite{Li:2009br,nonpert}. Nevertheless, the anomalous
dimension of the soft function given below is expected to be free of
nonperturbative corrections, hence its evolution may still provides
valuable information. A close investigation of the soft function in
SCET is beyond the scope of the present paper, and we refer the
reader to Ref.~\cite{Li:2009br,nonpert} for the detailed discussion.

The evolution equation of the soft function can be derived from the
fact that the cross section in the threshold region is independent
of the factorization scale. In moment space the soft function obeys
the evolution equation
\begin{equation}
 \frac{d}{d\ln\mu}S_N(M^2_i,\mu)=\left(-2\gamma_1(\mu)+2\gamma_2(\mu)\right)
S_N(M^2_i,\mu),
\label{softad}
\end{equation}
where $\gamma_2(\mu)$ governs the DGLAP evolution for the PDFs~\cite{Altarelli:1977zs}
\begin{equation}
 \frac{d}{d\ln\mu}f^N_{qq}(x,\mu)
=-\gamma_2 (\mu) f^N_{qq}(x,\mu),
\end{equation}
and similarly for $f^N_{\bar{q}\bar{q}}(x,\mu)$. It can be
shown~\cite{Phys.Rev.D68.114019} to all orders in perturbation
theory that the anomalous dimension $\gamma_2(\mu)$ is a linear
function of $\ln \bar{N}$
\begin{equation}
 \gamma^N_2 (\mu)=\sum^{\infty}_{n=1} \left[\left(
\frac{\alpha_s}{\pi}\right)^n
\left( B^{(n)}_1\ln \bar{N} +B^{(n)}_0\right)\right],
\end{equation}
where the expanding coefficients $B^{(n)}_{1,0}$ can be obtained
from two loop splitting function~\cite{dglap}
\begin{eqnarray}
 B^{(1)}_1&=& 2 A^{(1)}_1, \qquad B^{(1)}_0=-\frac{3}{2}C_F,
\nonumber
\\
B^{(2)}_1 &=& 2A^{(2)}_1.
\end{eqnarray}
Note that at $\mathcal{O}(\alpha_s)$, the term $B^{(1)}_0$ in
$\gamma_2$, which leads to single logarithms evolution, coincides
with the corresponding term in $\gamma_1$. It can then be seen from
Eq.~(\ref{softad}) that the anomalous dimension of the soft function
doesn't lead to single logarithms evolution. However, this is no
longer true at the NNLO \cite{Becher:2007ty}. We also notice that
the peculiar $\ln\bar{N}$ evolution in the soft function anomalous
dimension corresponds to a plus distribution evolution in momentum
space. With the extra angular restrictions on the real gluon, it
originates from the incomplete cancellation between real and virtual
corrections in the threshold region.

Combining the above results, we can write down the resummed cross
section in moment space
\begin{eqnarray}
 \sigma^{\rm NLL}_N(M^2_i,S)
&=&\sigma_0\sum_{a,b}f^N_{a/A}(x_a,\mu_f)
f^N_{b/B}(x_b,\mu_f)|C_V(M^2_i,\mu_f)|^2 S_N(M^2_i,\mu_f) \nonumber
\\
\label{eq:scet} &=&\sigma_0\sum_{a,b}f^N_{a/A}(x_a,\mu_f)
f^N_{b/B}(x_b,\mu_f)|C_V(M^2_i,\mu_r)|^2 S_N(M^2_i,\mu_I) e^{I_1+I_2},
\end{eqnarray}
with
\begin{equation}
  \label{eq:int}
  I_1 = -2 \int_{\mu_I}^{\mu_r}
\frac{d\mu}{\mu} \gamma_1(\mu), \qquad
  I_2 = 2\int_{\mu_I}^{\mu_f} \frac{d\mu}{\mu} \gamma_2(\mu).
\end{equation}

Now we can evaluate the integrals in Eq.~(\ref{eq:int}) using the
two-loop evolution of $\alpha_s$ in the $\overline{\rm MS}$
scheme. Keeping only terms up to NLL in the exponents in Eq.~(\ref{eq:scet}),
we obtain
\begin{eqnarray}
  I_1 + I_2 &=&
 \ln{N} g_1(\beta_0\alpha_s(\mu_r)\ln{N}/\pi)
  + g_2(\beta_0\alpha_s(\mu_r)\ln{N}/\pi)
\nonumber
\\
&&+\mathcal{O}(\alpha_s(\alpha_s\ln{N})^k),
\end{eqnarray}
with
\begin{eqnarray}
  g_1(\lambda) &=& \frac{A_1^{(1)}}{\beta_0\lambda} \left[ 2\lambda + (1-2\lambda)
  \ln(1-2\lambda) \right],
  \\
  g_2(\lambda) &=& - \frac{2A^{(1)}_1\gamma_E}{\beta_0} \ln(1-2\lambda) +
  \frac{A^{(1)}_1\beta_1}{\beta_0^3} \left[ 2\lambda + \ln(1-2\lambda)
+ \frac{1}{2}\ln^2(1-2\lambda) \right]
\nonumber
\\
&& -\frac{A^{(2)}_1}{\beta_0^2} \left[
 2\lambda
  + \ln(1-2\lambda) \right]
- \frac{A^{(1)}_1}{\beta_0} \ln(1-2\lambda) \ln\frac{\mu_r^2}{M_i^2} -
  \frac{A^{(1)}_1}{\beta_0} 2\lambda \ln\frac{\mu_r^2}{\mu_f^2}
\nonumber
\\
&&   + \frac{A^{(1)}_0-B^{(1)}_0}{\beta_0} \ln(1-2\lambda),
\end{eqnarray}
where $\beta_0$ and $\beta_1$ are the first two coefficients of the
QCD $\beta$ function~\cite{Tarasov:1980au}:
\begin{equation}
 \beta_0=\frac{1}{12}(11C_A-2n_f),
\qquad
\beta_1=\frac{1}{24}(17C^2_A-5C_An_f-3C_Fn_f).
\end{equation}
The NLL cross section in moment space is then given by
\begin{equation}
  \sigma^{\rm NLL}_N = \sigma_0\sum_{a,b}f^N_{a/A}(x_a,\mu_f)
f^N_{b/B}(x_b,\mu_f)|C_V(M^2_i,\mu_r)|^2
S_N(M^2_i,\mu_I)
\exp(g_1\ln{N}+g_2).
\label{bbhnll}
\end{equation}
Note that the NLL resummed cross section for SM Higgs boson
production can be obtained from Eq.~(\ref{bbhnll}) by setting
$\kappa_i=1$ and $\Delta^{\rm SUSY}=0$. To obtain the physical cross
section, we perform the inverse Mellin transformation back to the
$x$-space
\begin{equation}
  \sigma^{\text{NLL}}(\tau) = \frac{1}{2\pi i} \int_C dN \tau^{-N} \sigma^{\text{NLL}}_N.
\end{equation}
Here the integral contour is chosen as the minimal
prescription~\cite{Catani:1996yz} and the tricks introduced in
Ref.~\cite{Phys.Rev.D66.014011} is used to evaluate the $N$-integral
numerically. Note that in the minimal prescription, the Landau
pole singularities in the numerical inverse Mellin transformation is avoided
by choosing the contour that doesn't include this pole.
Another approach developed recently is to solve the RG equation
in momentum space
directly~\cite{Becher:2007ty}, which we do not consider here.
 Finally, The resummed cross section at NLL accuracy is
defined to be the NLL cross section plus the remaining terms in the
NLO result which are not resummed, i.e.,
\begin{equation*}
  \sigma^{\text{RES}} = \sigma^{\text{NLL}} + \sigma^{\text{NLO}} -
  \sigma^{\text{NLL}} \bigg|_{\alpha_s=0} - \alpha_s \left(
    \frac{\partial\sigma^{\text{NLL}}}{\partial\alpha_s} \right)_{\alpha_s=0}.
\end{equation*}

\section{NUMERICAL RESULTS AND DISCUSSION}

In this section, we present the numerical results for inclusive production
cross section of neutral
Higgs bosons at the LHC. In our numerical calculations the
following SM input parameters were chosen~\cite{Amsler:2008zz}:
\begin{equation}
\begin{array}{lll}
M_t=172.4\text{~GeV},& G_F=1.16637\times
10^{-5}\text{~GeV}^{-2},&
\alphas(M_Z)=0.1176, \\ M_W=80.398\text{~GeV}, &
M_Z=91.1876\text{~GeV}. & \\
\end{array}
\end{equation}
The running QCD coupling $\alpha_s$ was evaluated at the
two-loop level~\cite{runningalphas}, and the CTEQ6.6M
PDFs~\cite{Nadolsky:2008zw} were
used to calculate the various cross sections.
Moreover, in order to improve the perturbative calculations, 1-loop
and 2-loop running masses $m_b(\mu_r)$ are taken as
following~\cite{Braaten:1980yq,runningmb}:
\begin{equation}
\overline{m}_b (\mu_r)_{\rm 1-loop} = m^{\rm pole}_b \left(
\frac{\alpha_s(\mu_r)} {\alpha_s(m^{\rm pole}_b)}\right)^{c_0/b_0}
\end{equation}
for LO cross sections and
\begin{equation}
 \overline{m}_b (\mu_r)_{\rm 2-loop}=
m^{\rm pole}_b \left( \frac{\alpha_s(\mu_r)} {\alpha_s(m^{\rm
pole}_b)}\right)^{c_0/b_0} \left[1+\frac{c_0}{\pi b_0}(c_1-b_1)
\left(\alpha_s(\mu_r)-\alpha_s(m^{\rm pole}_b)\right) \right]
\end{equation}
for NLO cross sections, respectively, where
\begin{equation}
\begin{array}{ll}
b_0=\frac{1}{4\pi}\left(\frac{11}{3}N_c
-\frac{2}{3}N_f \right),
&
b_1=\frac{1}{2\pi}\left( \frac{51 N_c-19 N_f}
{11 N_c-2 N_f} \right),
\\
c_0=\frac{1}{\pi}, & c_1=\frac{1}{72\pi}(101N_c-10N_f).

\end{array}
\end{equation}
In addition, to resum the leading $\tan\beta$ enhanced effects from
SUSY QCD corrections, the LO cross sections $\sigma_0$ is replaced
by the ``Improved Born
Approximation''(IBA)~\cite{runningmb,Guasch:2003cv,Dittmaier:2006cz}
as following:
\begin{eqnarray}
\sigma^h_{\rm IBA}&=&
\frac{{\sigma}^h_0}{(1+\Delta_b)^2}
\left(1-\Delta_b\frac{1}{\tan\alpha\tan\beta}\right)^2,
\nonumber
\\
{\sigma}^H_{\rm IBA}&=& \frac{{\sigma}^H_0}{(1+\Delta_b)^2}
\left(1+\Delta_b\frac{\tan\alpha}{\tan\beta}\right)^2,
\nonumber
\\
{\sigma}^A_{\rm IBA}&=& \frac{{\sigma}^A_0}{(1+\Delta_b)^2}
\left(1-\frac{\Delta_b}{\tan^2\beta}\right)^2,
\end{eqnarray}
with
\begin{equation}
\Delta_b
=
\frac{\alpha_s}{2\pi}
C_Fm_{\tilde{g}}\mu\tan\beta I(m_{\tilde{b}_1},
m_{\tilde{b}_2},m_{\tilde{g}}),
\end{equation}
where $I(a,b,c)$ is defined as
\begin{eqnarray}
I(a,b,c)=\frac{-1}{(a^2-b^2)(b^2-c^2)(c^2-a^2)}
(a^2b^2\ln\frac{a^2}{b^2} +b^2c^2\ln\frac{b^2}{c^2}
+c^2a^2\ln\frac{c^2}{a^2}).
\end{eqnarray}
To avoid double counting, it is necessary to subtract the
corresponding SUSY QCD corrections from the renormalization constant
$\delta m_b$ in the following numerical calculations.

All the MSSM parameters are generated with
FeynHiggs~\cite{Heinemeyer:1998yj}. For simplicity, we only present
numerical results for the $m^{\rm max}_h$ scenario, which is
suitable for the MSSM Higgs boson search at hadron
colliders~\cite{Carena:2002qg}, and the resummation effects on total
cross sections for other scenarios are almost the same. In the
$m^{\rm max}_h$ scenario~\cite{Heinemeyer:1999zf}, The parameters
are:
\begin{eqnarray}
 &&
M_{\rm SUSY}=1\ {\rm TeV},\quad \mu=200\ {\rm GeV},
\quad M_2=200\ {\rm GeV},
\nonumber
\\
&& X_t=2M_{\rm SUSY},\quad A_b=A_t,\quad m_{\tilde{g}}=0.8M_{\rm
SUSY}, \label{mhm}
\end{eqnarray}
where $X_t=A_t-\mu\cot\beta$ and $M_2$ is the wino mass term.

Moreover, to show the resummation effects in SM (MSSM) Higgs boson
production, we define
\begin{equation}
 \delta K=\frac{\sigma^{\rm RES}-\sigma^{\rm NLO}}
{\sigma^{\rm NLO}},
\end{equation}
where $\sigma^{\rm NLO}$ includes NLO QCD and SUSY QCD corrections.

In Fig.~\ref{smcompare} we show the K factor, which is defined as
the ratios between the cross sections at the higher orders and the
cross sections at the LO, for SM Higgs production cross sections
with the NLL resummation effects and the NNNLO collinear and soft
gluon effects in Ref.~\cite{Kidonakis:2007ww}, respectively,
assuming $\mu_r=\mu_f=M_h$. We can see that our result is about
$5$\%, while their result shown in Ref.~\cite{Kidonakis:2007ww} is
about $35$\%.

Fig.~\ref{smmass} shows the total cross sections for SM Higgs
production as functions of the mass of Higgs boson including higher
order QCD effects, assuming $\mu_r=\mu_f=M_h$. It can be seen from
the figure that the total cross sections become small as the Higgs
boson mass increases, which is due to the decreasing of the bottom
quark density. The figure shows that the threshold resummation
effects reduce the LO results significantly, and enhance the NLO
results by a few percent generally.

Fig.~\ref{smrenscale} shows the renormalization scale dependence of
SM Higgs production, assuming $\mu_f=M_h/4$. We find that
the renormalization scale dependence is reduced by the resummation
effects, and the resummed cross section is very close to the NLO
cross section in the vicinity of $\mu_r\sim M_h$.
If restricting the renormalization scale to a factor of
five above or below $M_h$, the scale dependence is reduced from
about 50\% at LO to 38\% at NLO, and to about 30\% at NLL.

Fig.~\ref{smfacscale} gives the factorization scale dependence of SM
Higgs production, assuming $\mu_r=M_h$. We see that the NLO
corrections reduces the scale dependence significantly, and the
resummation effects can not improve the scale dependence. This is
due to the fact that the dominate contribution to the reduction of
factorization scale dependence comes from the splitting processes of
the initial states at the NLO, which are not included in the NLL
resummation effects, while the NNLO
corrections~\cite{Harlander:2003ai} can further reduce the
factorization scale dependence. As shown in the figure, the
resummation effects is small around $\mu_f\sim M_h/4$, which implies
that the convergence of the resummed logarithmic terms is very well around
such scale.

In Fig.~\ref{smK} the resummation effects $\delta K$ are presented
as a function of the SM Higgs boson mass $M_h$, assuming $\mu_r=M_h$
for $\mu_f=M_h/4$ and $\mu_f=M_h$, respectively. The results show
that the resummation effects are quite small for $\mu_f=M_h/4$,
about -1\%, but for $\mu_f=M_h$, $\delta K$ can be over 5\%. The
resummation effects do not lead to large corrections relative to the
NLO results, since the scaling variable $\tau=M^2_h/s$ is far away
from 1 and the falling off of the bottom quark density is smooth.

Figs. 7-12 show the total cross sections for $h$, $H$ and $A$
production as functions of their masses for the $m^{\rm max}_h$
scenario, assuming $\tan\beta=5$ and $30$, respectively.
In these figures and the following, the variation on $M_h$ and
$M_H$ is obtained from varying $M_A$. We find
that in all cases, the NLO SUSY QCD corrections are negligible
after employing the Improved Born Approximation, and the threshold
resummation cross sections reduce the LO cross sections and enhance
the NLO cross sections by a few percent.

Figs.~\ref{mhmmhK} and \ref{mhmmhtb30K} present the resummation
effects $\delta K$ as a function of $M_h$ assuming $\mu_r=M_h$ for
$\tan\beta=5$ and $30$, respectively. In general, $\delta K$ is
about -1\% when $\mu_f=M_h/4$. However, the $\delta K$ can be larger
than 5\% when $\mu_f=M_h$. Note that the SUSY QCD corrections has
little impact on the resummation effects $\delta K$ as shown in the
figures, since the SUSY QCD do not give rise to new soft gluon
interaction.

\section{CONCLUSION}

In conclusion, we have calculated the QCD effects in the production
of the neutral Higgs boson via bottom quark fusion in both the SM
and the MSSM at the LHC, which include not only the NLO QCD and SUSY
QCD corrections, but also the NLL threshold resummation effects in
the framework of SCET. Similar to Drell-Yan
production~\cite{Becher:2007ty,drellyan}, The resummation is
achieved by separating the contribution from hard and soft scale
into different matching coefficients and then summing the large
logarithms of scale ratio via RG equation. This approach has the
advantage that the ambiguity due to the Landau pole singularities is
reduced. Our results of the NLO QCD and SUSY QCD corrections agree
with the calculations reported in the previous literatures, and the
resummatioin effects are about -1\% and 5\% for $\mu_f=M_h/4$ and
$\mu_f=M_h$, respectively.

\begin{acknowledgments}
This work was supported in
part by the National Natural Science Foundation of China, under
Grants No.10721063 and No.10635030.
\end{acknowledgments}

\appendix

\section{}
In this appendix, we collect the relevant MSSM Feynman rules
as following~\cite{higgshunter,Kraml:1999qd}:

1. The coupling between neutral Higgs boson and
sbottom $H_k-\tilde{b}_l-{\tilde{b}_m}^*$:
\begin{equation*}
i[R^{\tilde{b}}\hat{G}^{(k)}(R^{\tilde{b}})^T]_{lm}=
i(G^{(k)})_{lm},
\end{equation*}
with
\begin{eqnarray*}
\hat{G}^{(h)}&=& \frac{g_2}{M_W}
\left(\begin{array}{cc}
M^2_Z \sin (\alpha+\beta)C_{bL}+
m^2_b\frac{\sin\alpha}{\cos\beta}
&
\frac{m_b}{2} \left(
A_b \frac{\sin\alpha}{\cos\beta}+\mu\frac{\cos\alpha}
{\cos\beta}\right)
\\
\frac{m_b}{2} \left(
A_b \frac{\sin\alpha}{\cos\beta}+\mu\frac{\cos\alpha}
{\cos\beta}\right)
&
-M^2_Z\sin (\alpha+\beta) C_{bR}
+m^2_b\frac{\sin\alpha}{\cos\beta}
\end{array} \right),
\end{eqnarray*}
\begin{eqnarray*}
\hat{G}^{(H)}=\frac{g_2}{M_W}
\left(\begin{array}{cc}
-M^2_Z \cos (\alpha+\beta) C_{bL}
-m^2_b\frac{\cos\alpha}{\cos\beta}
 &
-\frac{m_b}{2} \left(
A_b\frac{\cos\alpha}{\cos\beta}
-\mu\frac{\sin\alpha}{\cos\beta}\right)
\\
-\frac{m_b}{2} \left(
A_b\frac{\cos\alpha}{\cos\beta}
-\mu\frac{\sin\alpha}{\cos\beta}\right)
&
M^2_Z \cos (\alpha+\beta) C_{bR}
-m^2_b \frac{\cos\alpha}{\cos\beta}
 \end{array} \right),
\end{eqnarray*}
\begin{eqnarray*}
\hat{G}^{(A)}=i\frac{g_2 m_b}{2m_W} \left(\begin{array}{cc} 0
& -A_b \tan\beta  -\mu \\
A_b \tan\beta +\mu & 0 \end{array} \right).
\end{eqnarray*}

2. The coupling between bottom, sbottom and gluino
$b-\tilde{b}^{(*)}_j-\tilde{g}$:

\bigskip

\begin{picture}(80,25)(-40,50)
 \put(25,0){\mbox{\epsfig{figure=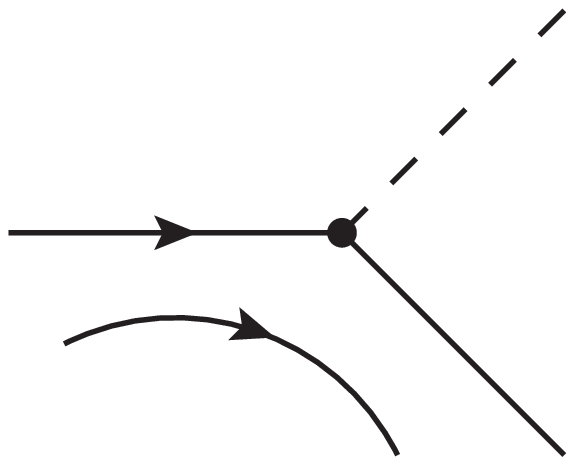,height=2.3cm}}}
\put(19,28){
\makebox(0,0)[br]{{\small $b$}}}
\put(110,-2.0){
\makebox(0,0)[bl]{{\small $\tilde{g}$}}}
\put(110,60.5){
\makebox(0,0)[bl]{{\small $\tilde{b}^*_j$}}}
\put(35,37){\makebox(0,0)[bl]{{\scriptsize $l$}}}
\put(85,59){\makebox(0,0)[tl]{{\scriptsize $k$}}}
\put(100,13){
\makebox(0,0)[bl]{{\scriptsize $a$}}}
\put(140,28){
\makebox(0,0)[bl]
{
$ -\sqrt{2} ig_s \mathbf{T}^a_{kl}
(R^{\tilde{b}}_{j1} P_L-R^{\tilde{b}}_{j2}P_R)$
}}
\end{picture}
\\
\\
\\
\bigskip

\begin{picture}(80,25)(-40,50)
 \put(25,0){\mbox{\epsfig{figure=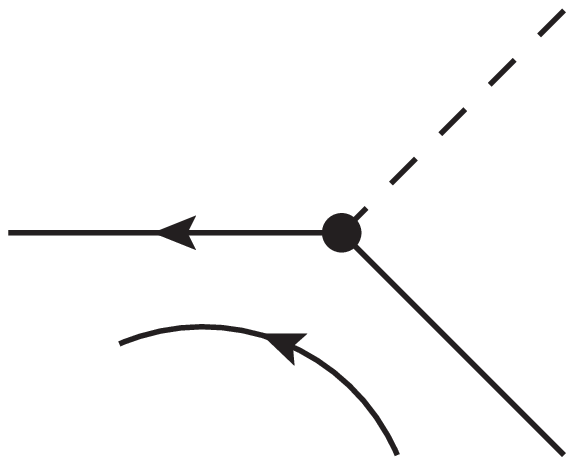,height=2.3cm}}}
\put(19,28){
\makebox(0,0)[br]{{\small $b$}}}
\put(110,-2.0){
\makebox(0,0)[bl]{{\small $\tilde{g}$}}}
\put(110,60.5){
\makebox(0,0)[bl]{{\small $\tilde{b}_j$}}}
\put(35,37){\makebox(0,0)[bl]{{\scriptsize $l$}}}
\put(85,59){\makebox(0,0)[tl]{{\scriptsize $k$}}}
\put(100,13){
\makebox(0,0)[bl]{{\scriptsize $a$}}}
\put(140,28){
\makebox(0,0)[bl]
{
$ -\sqrt{2} ig_s \mathbf{T}^a_{lk}
(R^{\tilde{b}}_{j1} P_R-R^{\tilde{b}}_{j2}P_L)$
}}
\end{picture}\\
\\
\\
\\
Here $P_{L(R)}=(1\mp\gamma_5)/2$ and $\mathbf{T}^a$ is the $SU(3)$
generator in fundamental representation.

\bibliography{bbhiggs}

\newpage

\begin{figure}[!hp]
\includegraphics[width=0.8\textwidth]{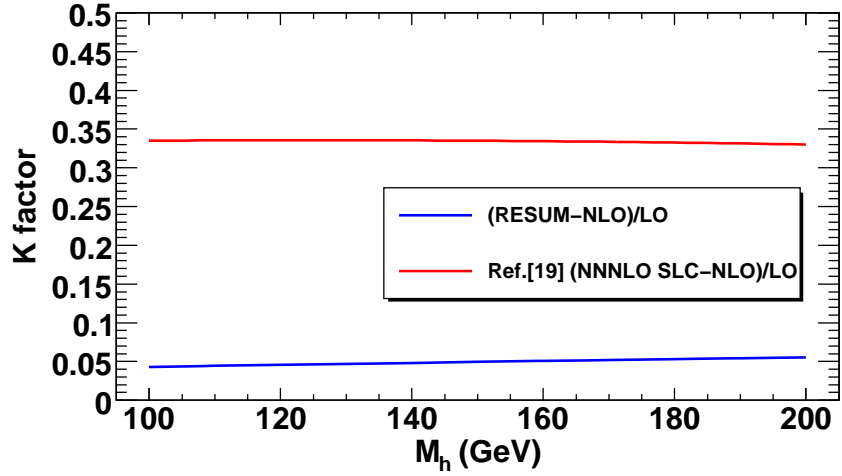}
\caption{The K factor of our results compared with those of Ref
\cite{Kidonakis:2007ww}, assuming $\mu_r=\mu_f=M_h$.}
\label{smcompare}
\end{figure}

\begin{figure}[!hp]
\includegraphics[width=0.8\textwidth]{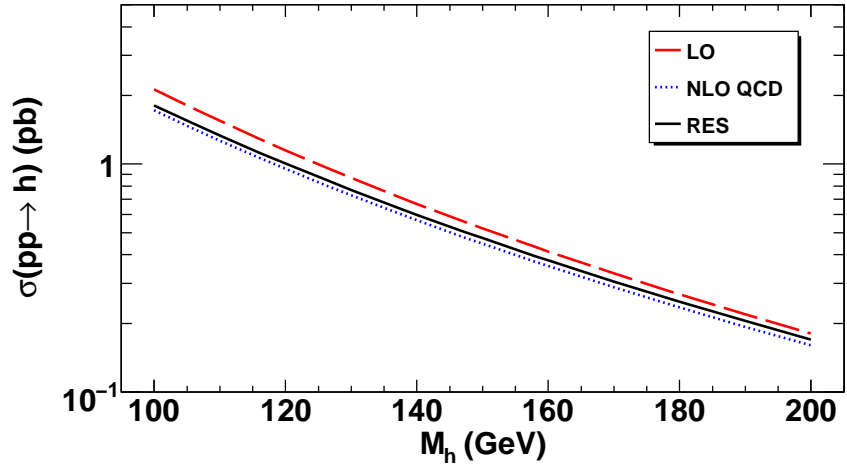}
\caption{The total cross section for $pp\rightarrow h+X$ in SM at
$\sqrt{s}=14$ TeV, assuming $\mu_r=\mu_f=M_h$.} \label{smmass}
\end{figure}

\begin{figure}[!hp]
\includegraphics[width=0.8\textwidth]{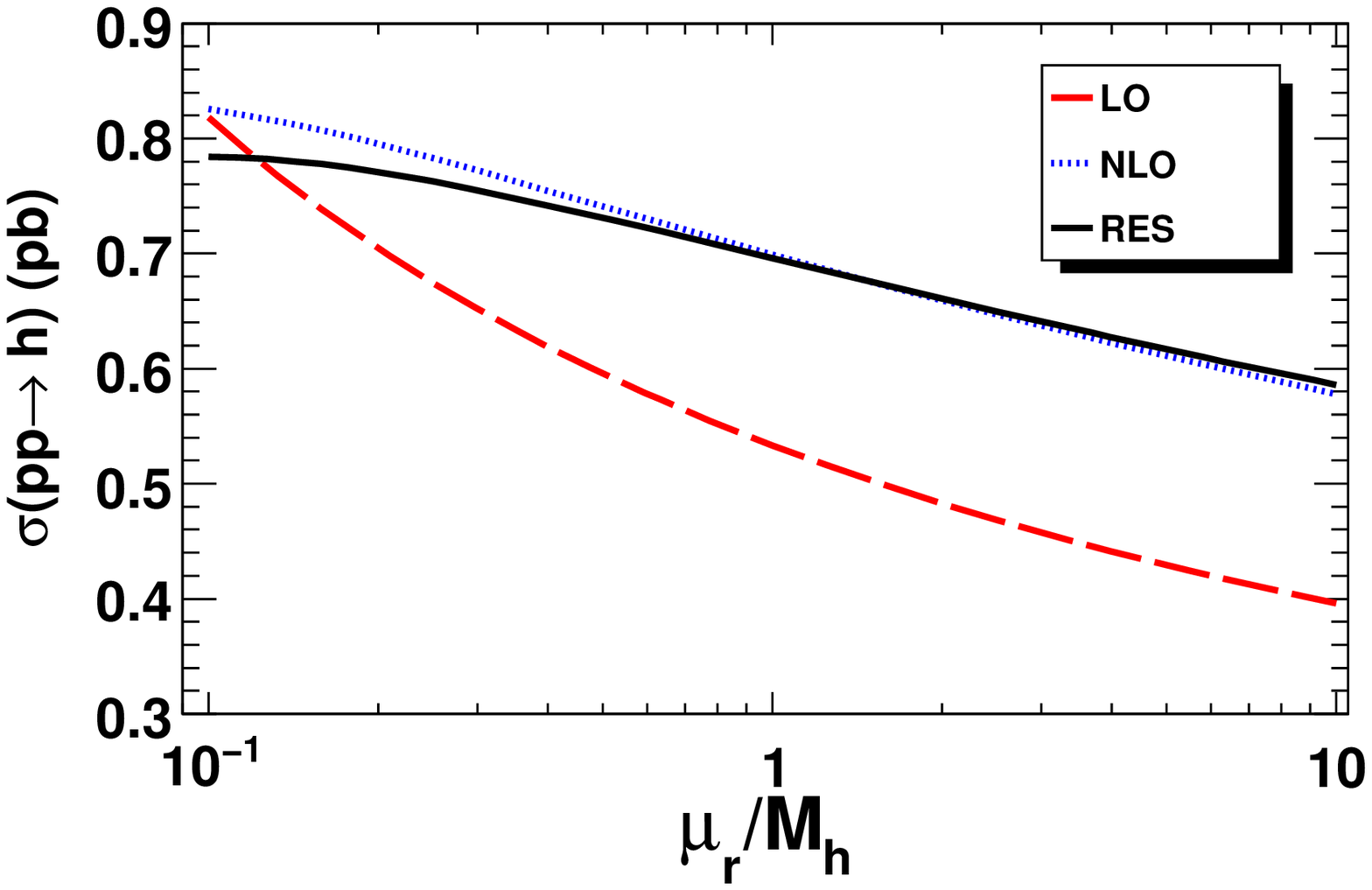}
\caption{The renormalization scale dependence of the total cross
section for $pp\rightarrow h+X$ in SM at $\sqrt{s}=14$ TeV, assuming
$\mu_f=M_h/4$ and $M_h=120$ GeV.} \label{smrenscale}
\end{figure}

\begin{figure}[!hp]
\includegraphics[width=0.8\textwidth]{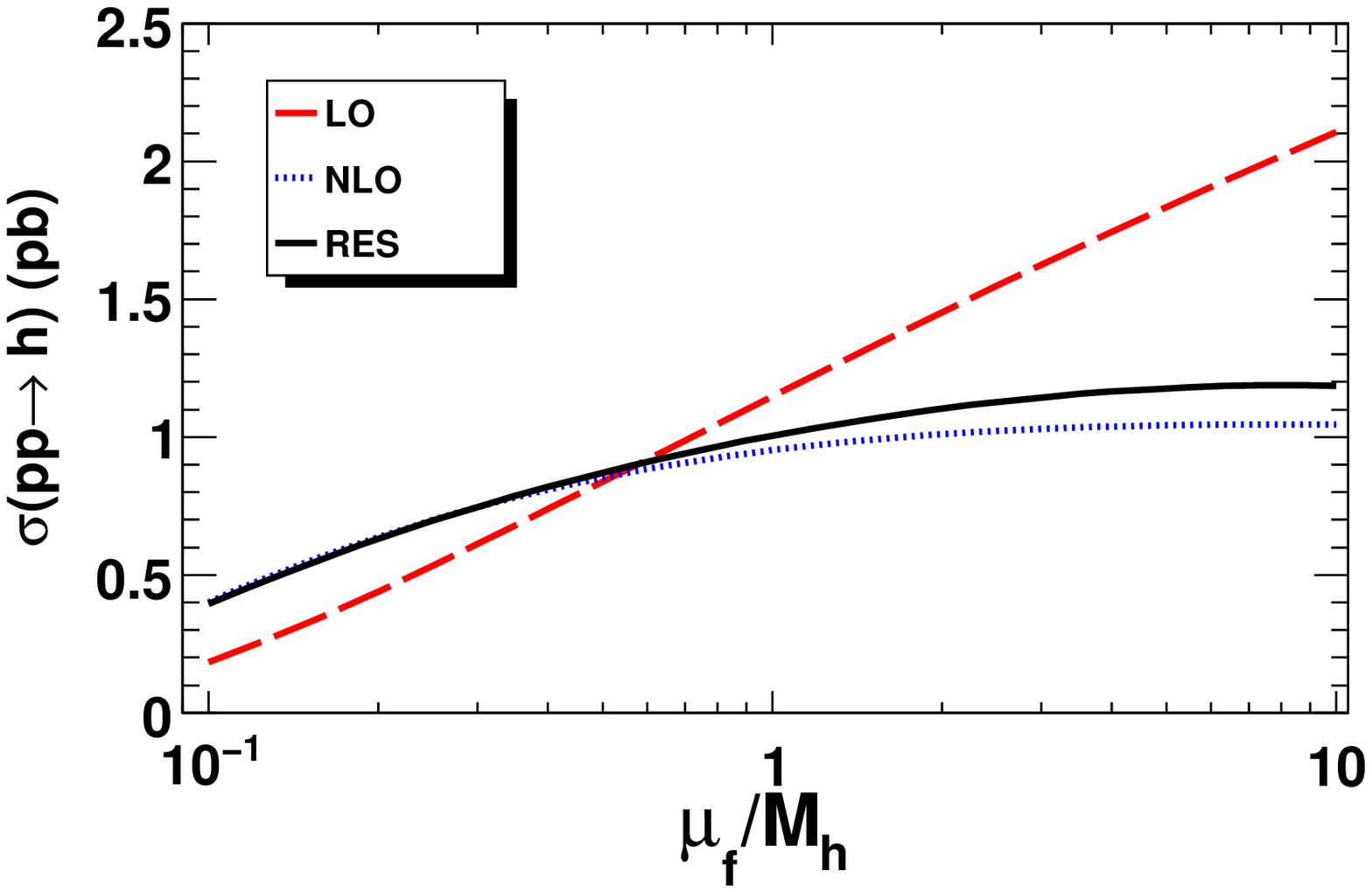}
\caption{The factorization scale dependence of the total cross
section for $pp\rightarrow h+X$ in SM at $\sqrt{s}=14$ TeV,
assuming $\mu_r=M_h$ and $M_h=120$ GeV.}
\label{smfacscale}
\end{figure}

\begin{figure}[!hp]
\includegraphics[width=0.8\textwidth]{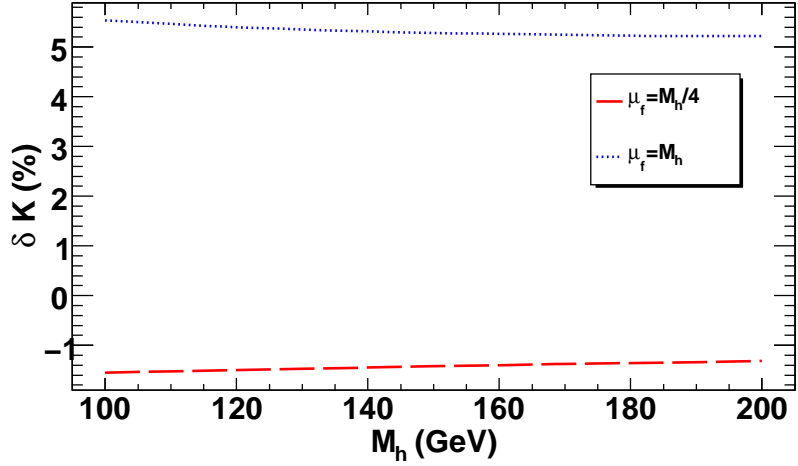}
\caption{The NLL resummation effects with different factorization
scale for $pp\rightarrow h+X$ in SM at $\sqrt{s}=14$ TeV, assuming
$\mu_r=M_h$.} \label{smK}
\end{figure}

\begin{figure}[!hp]
\includegraphics[width=0.8\textwidth]{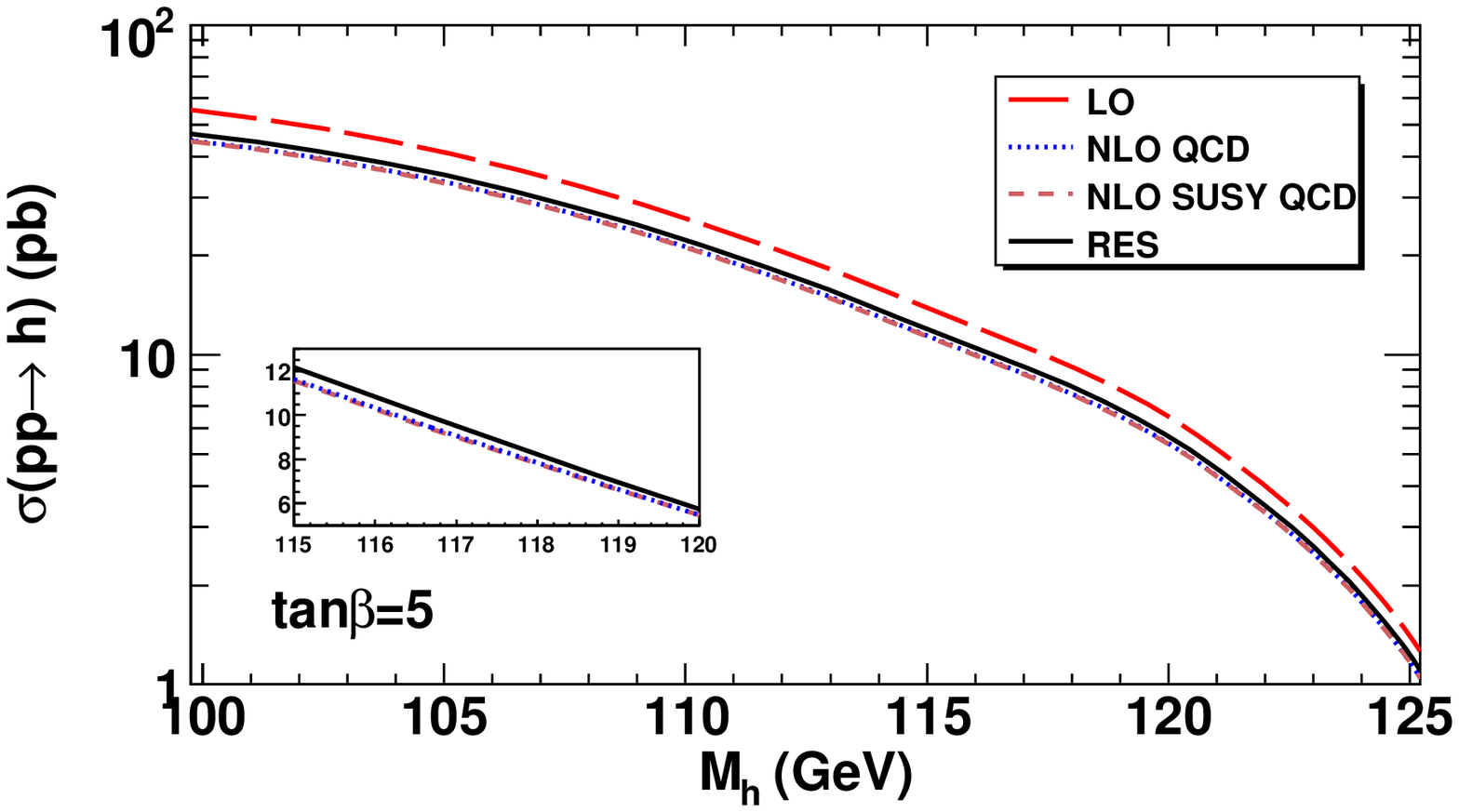}
\caption{The MSSM total cross section for $pp\rightarrow h +X$ at
$\sqrt{s}=14$ TeV in the $m^{\rm max}_h$ scenario,
assuming $\mu_r=\mu_f=M_h$ and $\tan\beta=5$.} \label{mhmmhmass}
\end{figure}

\begin{figure}[!hp]
\includegraphics[width=0.8\textwidth]{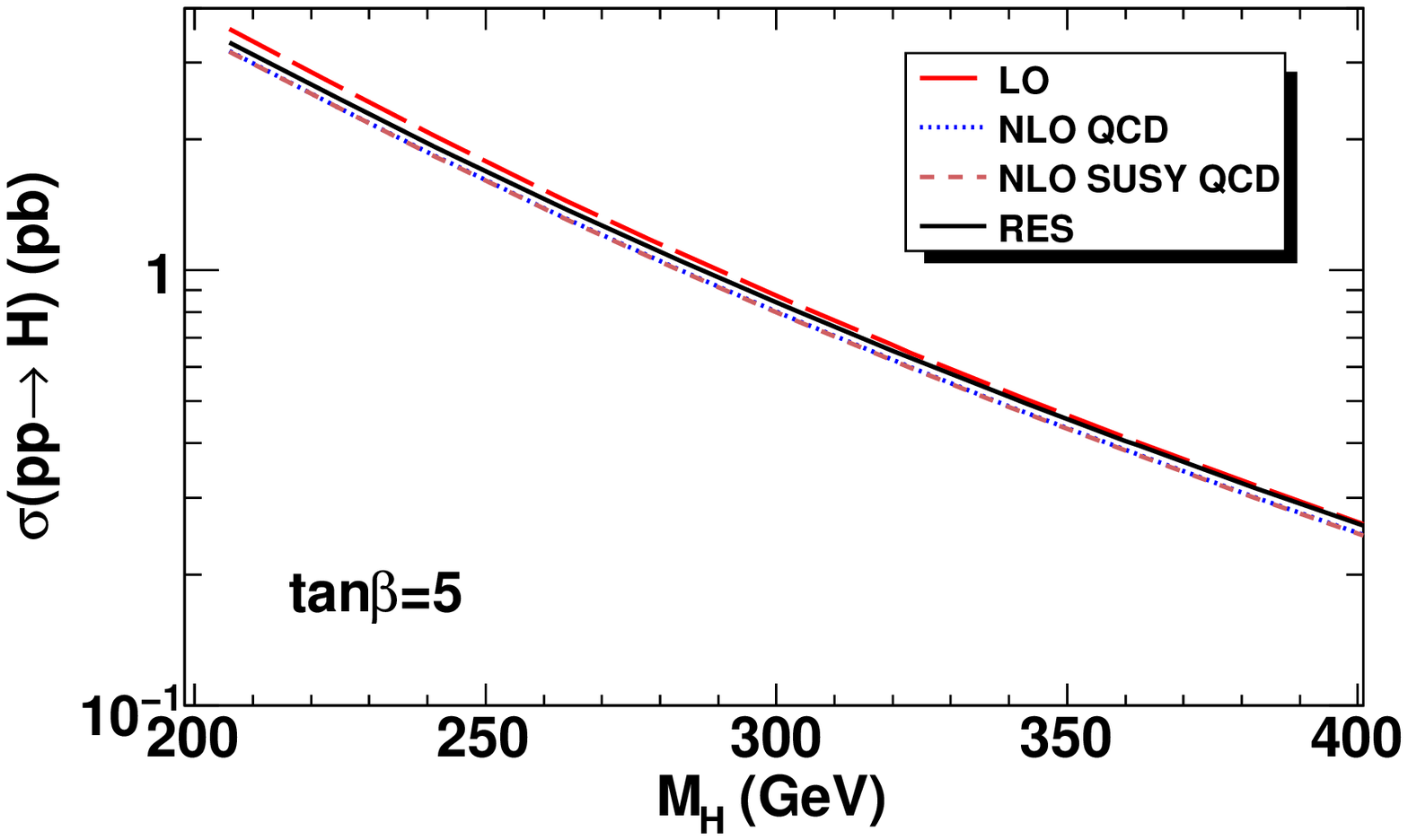}
\caption{The total cross section for $pp\rightarrow H +X$ at
$\sqrt{s}=14$ TeV in the $m^{\rm max}_h$ scenario,
assuming $\mu_r=\mu_f=M_H$ and $\tan\beta=5$.} \label{mhmmHmass}
\end{figure}

\begin{figure}[!hp]
\includegraphics[width=0.8\textwidth]{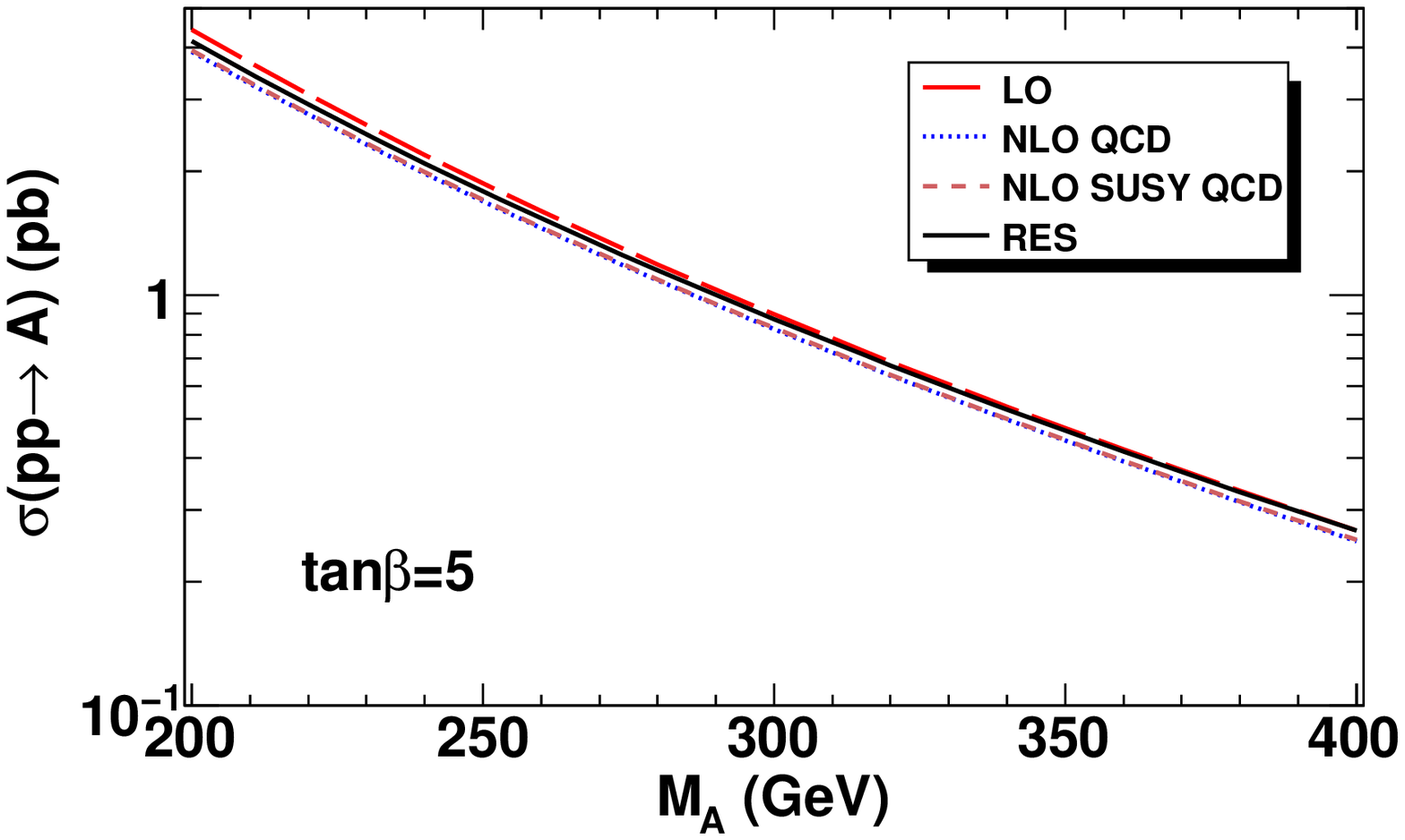}
\caption{The total cross section for $pp\rightarrow A +X$ at
$\sqrt{s}=14$ TeV in the $m^{\rm max}_h$ scenario,
assuming $\mu_r=\mu_f=M_A$ and $\tan\beta=5$.} \label{mhmmAmass}
\end{figure}

\begin{figure}[!hp]
\includegraphics[width=0.8\textwidth]{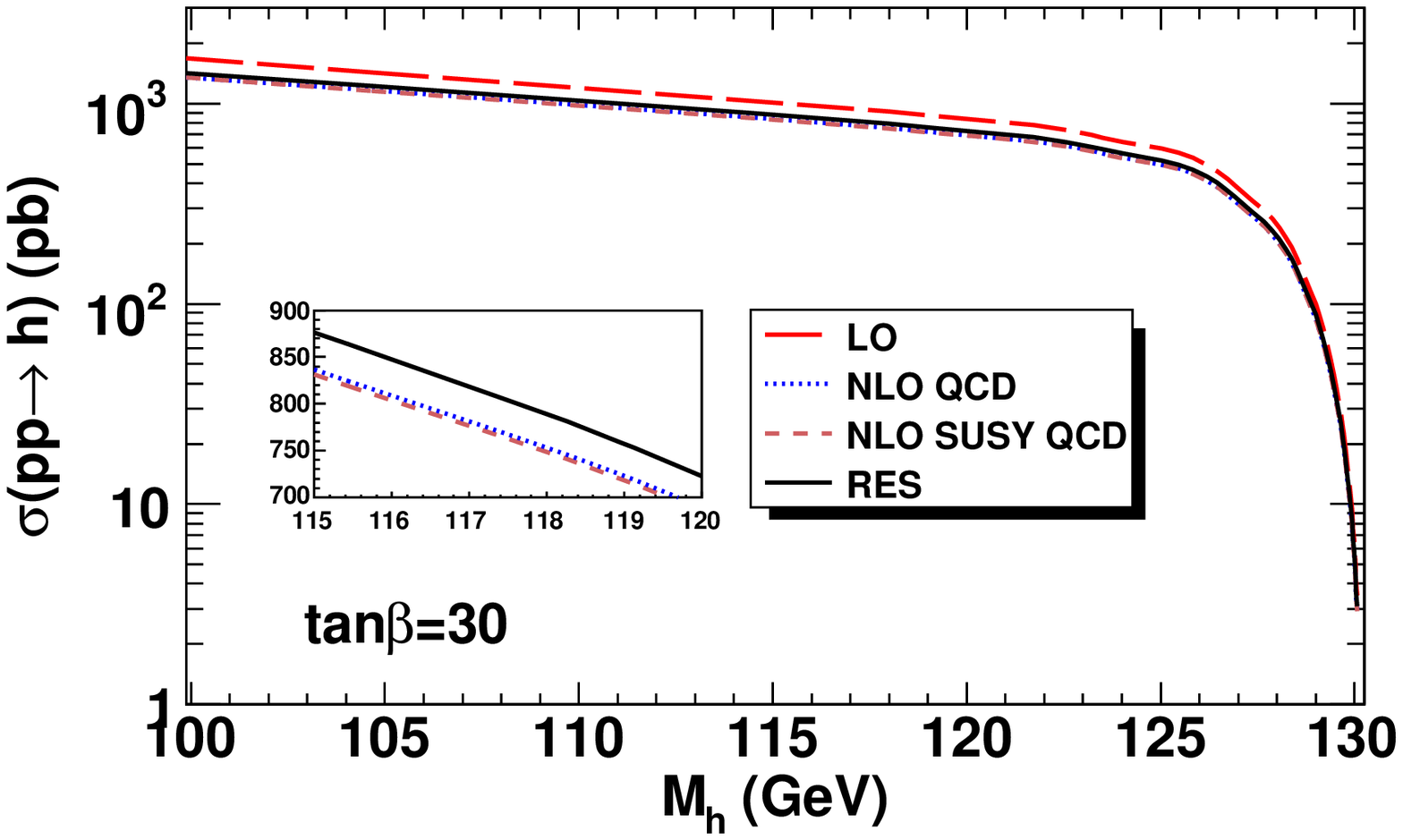}
\caption{The total cross section for $pp\rightarrow h +X$ at
$\sqrt{s}=14$ TeV in the $m^{\rm max}_h$ scenario,
assuming $\mu_r=\mu_f=M_h$ and $\tan\beta=30$.} \label{mhmmhmasstb30}
\end{figure}

\begin{figure}[!hp]
\includegraphics[width=0.8\textwidth]{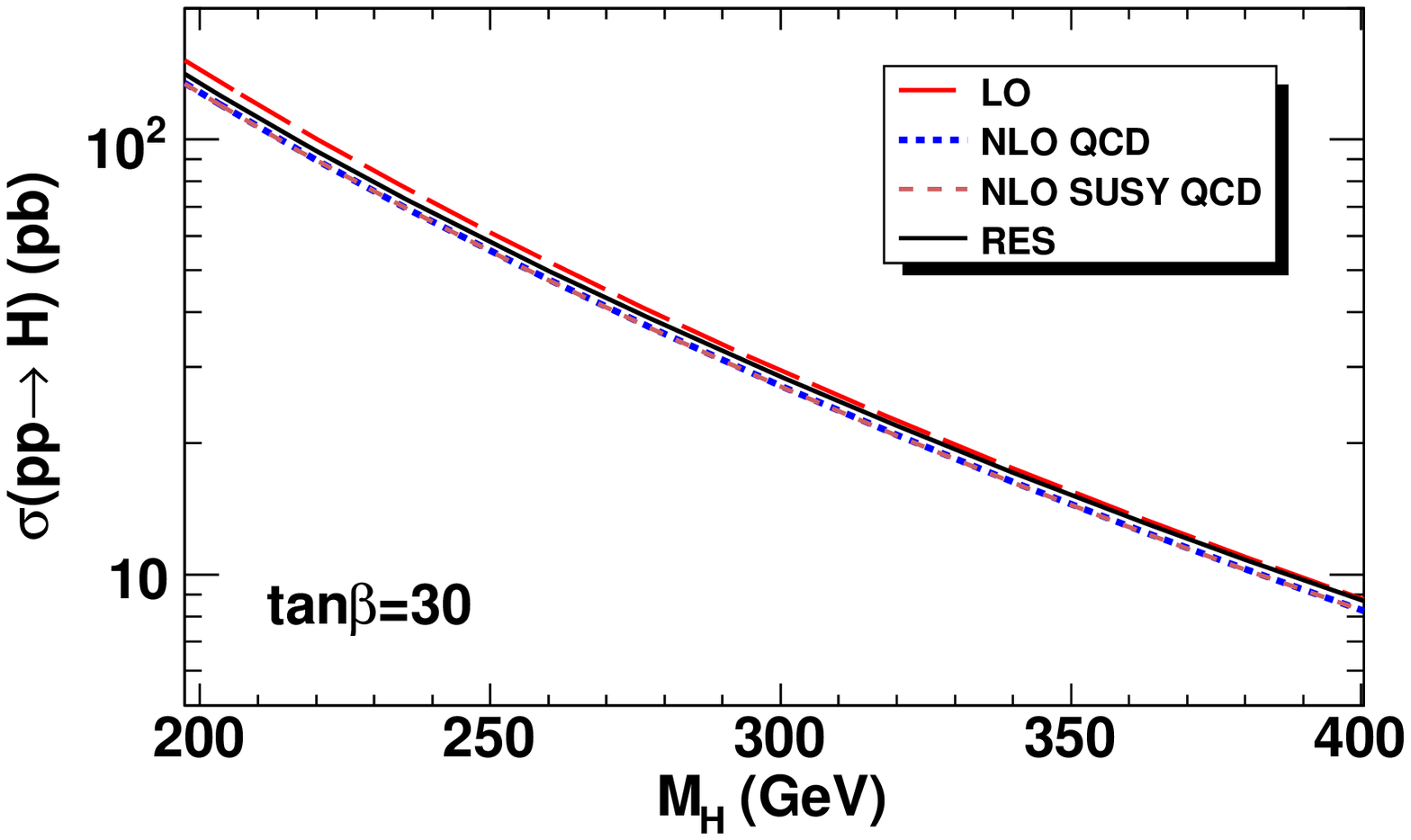}
\caption{The total cross section for $pp\rightarrow H +X$ at
$\sqrt{s}=14$ TeV in the $m^{\rm max}_h$ scenario,
assuming $\mu_r=\mu_f=M_H$ and $\tan\beta=30$.} \label{mhmmHmasstb30}
\end{figure}

\begin{figure}[!hp]
\includegraphics[width=0.8\textwidth]{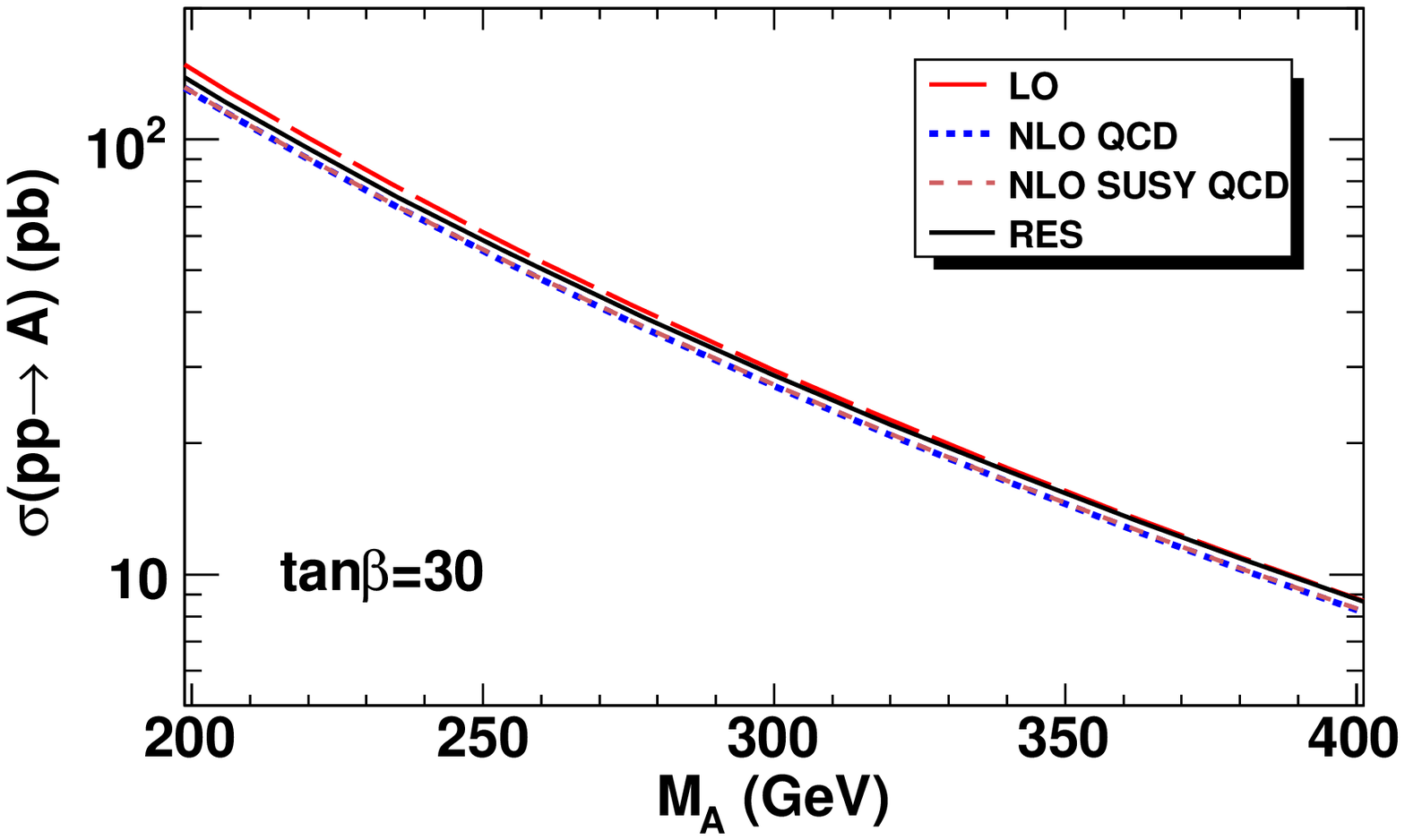}
\caption{The total cross section for $pp\rightarrow A +X$ at
$\sqrt{s}=14$ TeV in the $m^{\rm max}_h$ scenario,
assuming $\mu_r=\mu_f=M_A$ and $\tan\beta=30$.} \label{mhmmAmasstb30}
\end{figure}

\begin{figure}[!hp]
\includegraphics[width=0.8\textwidth]{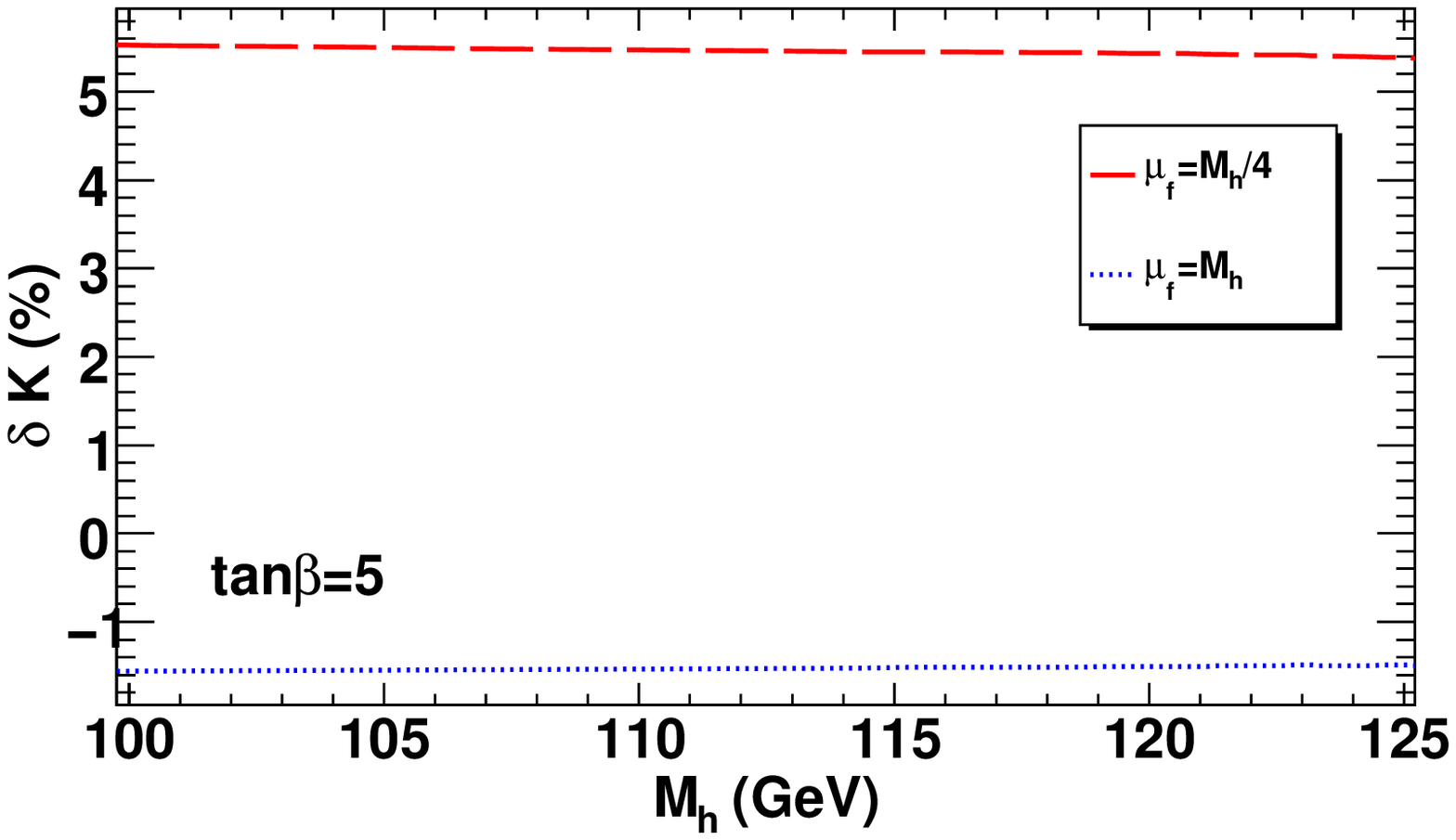}
\caption{The comparison of NLL resummation effects at different
factorization scale for $pp\rightarrow h +X$ at $\sqrt{s}=14$ TeV in
the $m^{\rm max}_h$ scenario, assuming
$\mu_r=M_h$ and $\tan\beta=5$.} \label{mhmmhK}
\end{figure}

\begin{figure}[!hp]
\includegraphics[width=0.8\textwidth]{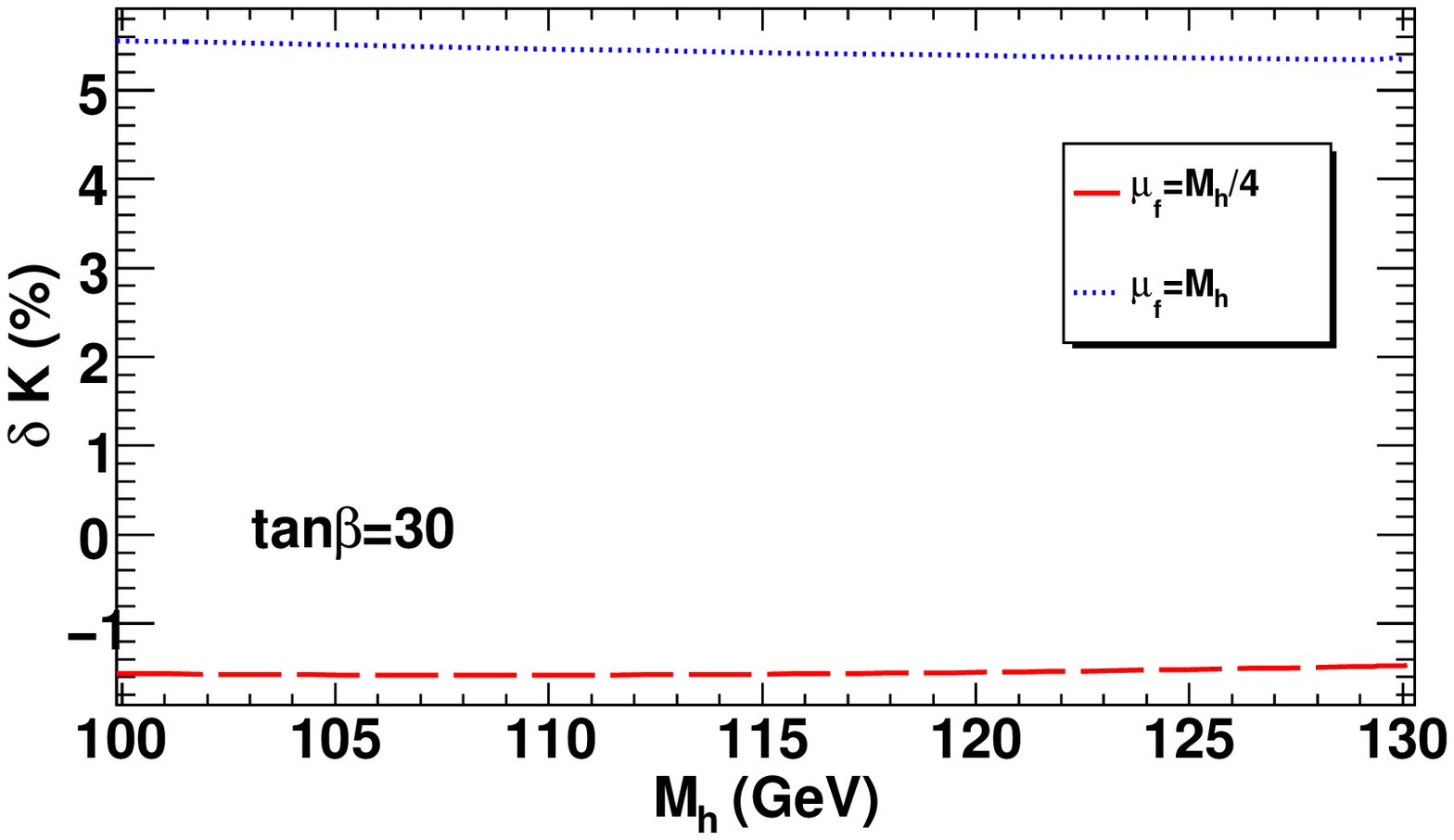}
\caption{The comparison of NLL resummation effects at different
factorization scale for $pp\rightarrow h +X$ at $\sqrt{s}=14$ TeV in
the $m^{\rm max}_h$ scenario, assuming
$\mu_r=M_h$ and $\tan\beta=30$.} \label{mhmmhtb30K}
\end{figure}

\end{document}